\documentclass[showpacs,showkeys,amsmath, amssymb]{ar-1col}

\usepackage{booktabs}
\usepackage{amsmath}
\usepackage{amssymb}
\usepackage{mathbbol}
\usepackage{graphicx}
\usepackage{dcolumn}
\usepackage{bm}
\usepackage[colorlinks=true, allcolors=blue]{hyperref}
\usepackage{soul}  

\usepackage{subfigure}
\graphicspath{{figures/}}
\usepackage[separate-uncertainty=false]{siunitx}
\usepackage{comment}
\usepackage{xspace}

\newcommand{\V}[1]{\mathbf{#1}}
\newcommand{\vr}{\mathbf{r}}
\newcommand{\vk}{\mathbf{k}}
\newcommand{\D}{\mathrm{div}}

\usepackage{xstring}

\newcommand*{\aref}[1]{%
	\IfBeginWith{#1}{eq:}{Eq.~\eqref{#1}}{}%
	\IfBeginWith{#1}{fig:}{Fig.~\ref{#1}}{}%
	\IfBeginWith{#1}{tab:}{Table~\ref{#1}}{}%
	\IfBeginWith{#1}{appendix:}{Appendix~\ref{#1}}{}%
	\IfBeginWith{#1}{sec:}{Section~\ref{#1}}{}%
}
\renewcommand{\vr}{\mathbf{r}}

\begin{document}
	
\title{Coherently Coupled  Mixtures of  Bose-Einstein Condensed Gases}

\author{Alessio Recati$^{1,2}$ and Sandro Stringari$^{1,2}$
\affil{$^1$ INO-CNR BEC Center, Dipartimento di Fisica, Universit\`a di Trento, 38123 Povo, Italy\\
$^2$Trento Institute for Fundamental Physics and Applications, INFN, 38123 Povo, Italy}}
	
\begin{abstract}
This paper summarizes some of the relevant features exhibited by binary mixtures of Bose-Einstein condensates in the presence of coherent coupling at zero temperature.  The coupling, which is experimentally produced by proper photon transitions, can either involve negligible momentum transfer from the electromagnetic radiation (Rabi coupling) or large momentum transfer (Raman coupling) associated with spin-orbit effects.  The nature of the quantum phases exhibited by coherently coupled  mixtures is discussed in detail,   including their paramagnetic, ferromagnetic, and, in the case of spin-orbit coupling, supersolid phases. The behavior of the corresponding elementary excitations is discussed, with explicit emphasis on the novel features caused by the spin-like degree of freedom.  Focus is further given to the topological excitations  (solitons, vortices) as well as to the superfluid properties. The paper also points out relevant open questions which deserve more systematic theoretical and experimental investigations.		
\end{abstract}
	
	\begin{keywords}
	Spinor Bose-Einstein condensates; Superfluidity; Magnetism; Spin-orbit coupling.  
	\end{keywords}
	
	\maketitle
	
	\section {Introduction}
	\label{sec:introduction}

Soon after the first experimental realization of Bose-Einstein condensation in cold gases of alkali atoms \cite{cornell1995,ketterle1995} the investigation of quantum mixtures \cite{Cornell97} has become a very popular subject of research in atomic physics, stimulating an extensive theoretical and experimental activity of both   fundamental and applicative interest. Important achievements have concerned, among others,  the realisation of novel quantum phases, 
the study   of collective dynamics and solitonic configurations, 
the  realization of polar molecules and the formation of  quantum droplets. First experiments
focused on alkali-metal gases , but there 
is currently a growing interest in mixtures composed of different atomic species, including Bose-Bose, Bose-Fermi and Fermi-Fermi mixtures. These studies, reported by an impressive  number of scientific works,  represent the  systematic implementation of the pioneering studies on  quantum degenerate mixtures  realized with helium  fluids \cite{Edwards1965,Baym1978}.

An intriguing possibility is given by the creation of coherent coupling among the different atomic species forming the mixture, giving rise to novel scenarios for non trivial equilibrium and non equilibrium many-body configurations. The aim of this paper is to summarize some of the most salient features exhibited by these configurations. For simplicity we will limit our discussion to the case of quantum mixtures occupying two different hyperfine states, hereafter called $|\uparrow\rangle$ and $|\downarrow\rangle$. Employing the usual spin $s=1/2$ representation these two single-particle states are classified as eigenstates of the  Pauli matrix operator $\sigma_z$ according to: $\sigma_z|\uparrow\rangle= +|\uparrow\rangle$ and $\sigma_z|\downarrow\rangle= -|\downarrow\rangle$.
The transfer of atoms between the two hyperfine states can be induced by proper photon transitions.  
Within the  rotating wave approximation, a suitable  polarization of the electromagnetic radiation,  the relevant single-particle spinor Hamiltonian takes the form:
\begin{equation}
h_{sp} = \frac{{\bf p}^2}{2m} - \frac{\hbar \Omega}{2}\sigma_x\cos (2k_0x - \Delta \omega_L t) - \frac{\hbar \Omega}{2}\sigma_y\sin (2k_0x - \Delta \omega_L t)+\frac{ \hbar \Delta \omega_{hf}}{2}\sigma_z
\label{HRaman}
\end{equation}
where ${\bf p}$ is the canonical momentum, $\Omega$ (herafter assumed real and positive) defines the intensity of the coupling of the atoms with the electromagnetic field, $k_0$ is the modulus of the wave vector difference  between the two  electromagnetic fields (hereafter chosen to be counter-propagating along the $x$-direction), and $\Delta\omega_L$ is the corresponding frequency difference. The energy $\hbar \Delta \omega_{hf}$ is the energy difference between the two hyperfine states, including the  non-linear  Zeeman effect. 

In the following we will distinguish the case when one can neglect the momentum transfer ($k_0=0$), which we will refer to as Rabi coupling, from the case where the value of $k_0$ cannot be ignored, which we will refer to as Raman (or spin-orbit) coupling.

The Hamiltonian (\ref{HRaman}) is not translational invariant, but exhibits a peculiar continuous screw-like symmetry, being invariant with respect to helicoidal translations of the form $\exp[id(p_x-\hbar k_0\sigma_z)/\hbar]$, consisting  of the  combination of a rigid translation with displacement $d$ and a spin rotation by the angle $-2dk_0$ around the $z$-axis. Translational invariance is obviously recovered for the Rabi coupling case.

The Hamiltonian (\ref{HRaman}) can be made time-independent and translational invariant by going to the so-called laser reference frame through the unitary transformation  
$U=\exp(i\Theta \sigma_z/2)$, corresponding to a position and time-dependent rotation in spin space by the angle $\Theta= 2k_0x-\Delta \omega_Lt$. The new Hamiltonian $h\rightarrow Uh_{sp}U^\dagger+i\hbar\dot{U}U^\dagger$ acquires the form
\begin{equation}
h_{\textrm{Rabi}} = \frac{{\mathbf{p}}^2}{2m} -\frac{\hbar \Omega}{2}\sigma _x + \frac{\hbar \delta}{2}\sigma_z
\label{hrf}
\end{equation}
\begin{equation}
h_{\textrm{SOC}} = \frac{1}{2m} \big[(p_x-\hbar k_0\sigma_z)^2 + p^2_\perp\big] +\frac{\hbar \Omega}{2}\sigma _x + \frac{\hbar \delta}{2}\sigma_z
\label{hSOC}
\end{equation}
where, for later convenience,  we have introduced the Rabi Hamiltonian $h_{\textrm{Rabi}}$ for $k_0=0$ and the spin-orbit Hamiltonian $h_{\textrm{SOC}}$ for $k_0\neq 0$ and the  detuning $\delta =\Delta \omega_L- \Delta\omega_{hf}$ is due to the additional time dependence of the unitary transformation. The spin-orbit term in Eq. (\ref{hSOC}) results from the non commutativity between the kinetic energy and the position dependent rotation. The spin-orbit Hamiltonian of Eq. (\ref{hSOC}) is characterized   by equal Rashba \cite{Rashba} and Dresselhaus \cite{dresselhaus} strengths. It is worth noticing that the canonical momentum 
$p_x=-i\hbar\partial_x$ entering the spin orbit Hamiltonian does not coincide with the physical momentum of particles, because of the presence of the spin term $\hbar k_0\sigma_z$. It is also useful to remark that the unitary transformation $U$ does not affect the density $n({\bf r})=n_\uparrow({\bf r})+ n_\downarrow({\bf r})$, nor  the $z$-component  $s_z({\bf r})=n_\uparrow({\bf r})- n_\downarrow({\bf r})$  of the  spin density. These quantities can be consequently safely calculated in the spin rotated frame, using the Hamiltonian (\ref{hSOC}).


In the following we will consider bosonic species which naturally undergo Bose-Einstein condensation at sufficiently low temperature and can reveal peculiar coherence effects associated with the Hamitonians discussed above. In the weakly interacting regime a 3D quantum mixture of bosonic atoms interacting with short range interactions is well described by 
mean-field (MF) theory, where the state
of the system is conveniently described by a two-component spinor wave function 	$\Psi({\bf r}, t)=(\Psi_\uparrow({\bf r,t}),\Psi_\downarrow({\bf r,t}))^T$,
normalized to the total number of particles 
$\int d{\bf r} \Psi^\dagger \Psi= N$
while the total energy of the system, including both
single-particle and interaction terms, can be expressed in terms of the relevant coupling constants as
\begin{equation}
E = \int d{\bf r}\varepsilon_{MF} = \int d{\bf r} [\Psi^{\dagger}h_{sp}\Psi + \frac{g_{dd}}{2}n^2 + \frac{g_{ss}}{2}s_z^2 ]
\label{Energy}
\end{equation}
where $n = \Psi^{\dagger}\Psi$ and $s_z= \Psi^{\dagger}\sigma_z\Psi$ are the total and spin density, respectively. 
The density-density and spin-spin coupling
constants are given by $g_{dd} = (g +g_{\uparrow \downarrow})/2$ and $g_{ss} =(g - g_{\uparrow \downarrow})/2= $, respectively and, for
simplicitly,  we have assumed   that   the
intraspecies couplings are equal, i.e., $g_{\uparrow \uparrow} = g_{\downarrow \downarrow}\equiv g>0$ and $g_{dd}>0$, to assure mean field stability against collapse. The most general case would include different 
intraspecies couplings. In this case 
one should replace $g$ with $(g_{\uparrow \uparrow} + g_{\downarrow \downarrow})/2$  and add a third term $g_{ds}ns_z$, with $g_{ds} =
(g_{\uparrow \uparrow} - g_{\downarrow \downarrow})/4$ inside the integral of Eq. (\ref{Energy}). The couplings
in each channel are related to the corresponding s-wave
scattering lengths via $g_{\sigma \sigma'}=4\pi \hbar^2 a_{\sigma \sigma'}/m$.

Starting from the energy functional (\ref{Energy}) one can derive coupled Gross-Pitaevskii equations for the separate components of the spinor wave function, whose predictions will be discussed in the next sections.

For reasons of space we are unable to discuss here  important results concerning coherently coupled mixtures where the mean-field description is not appropriate like, for example, the case of large enough attractive interspecies interaction, where beyond mean field   Lee-Huang-Yang corrections become crucially important  \cite{Bourdel-private} and even yield  self bound droplet states \cite{PetrovDroplet,Cappellaro2017,Sachdeva2020,Mazzanti2020}, and the case  of cold gases trapped in deep optical lattice where the system is properly described by Bose-Hubbard-like Hamiltonians allowing for strongly correlated configurations (see, e.g., \cite{Barbiero2016,Hofstetter2017,zhang2015,Hamner2015,Kartashov2016,Yamamoto2017}).

\section{Discrete and Continuous Symmetries}
\label{sec:symmetries}

The Hamiltonians discussed in the previous section exhibit important symmetries which are worth discussing because they permit to better understand the nature of the new equilibrium phases as well as the novel dynamic and superfluid features caused by  Rabi and spin-orbit coupling (SOC). 

Let us first consider the relevant discrete symmetries exhibited by our mixtures. $\mathbb{Z}_2$ is an important symmetry reflecting,   for vanishing detuning,  the symmetry of the Hamiltonian with respect to the exchange of the coordinates  of the two components. This symmetry is preserved in the presence of  the most relevant term proportional to $\Omega \sigma_x$. The spontaneous breaking of the $\mathbb{Z}_2$ symmetry is at the origin of the ferromagnetic phases exhibited both in the presence of Rabi and spin-orbit coupling, although the underlying mechanisms are different in the two cases (see Sect. \ref{sec:gsRabi}  and Sect. \ref{sec:phasesSOC}). Another important case of  discrete symmetries concerns time reversal and parity. They are both violated by the  spin orbit Hamiltonian (\ref{hSOC}) with crucial consequences on the breaking of the symmetry property $\omega(q) =\omega(-q)$ usually exhibited by the spectrum of the elementary excitations (see Sect.\ref{sec:excitationsSOC}). 

An important consequence of the term proportional to $\Omega$ in the Hamiltonian   is the violation of the continuous symmetry with respect to the relative phase $\phi_r= \phi_\uparrow - \phi_\downarrow$ of  the order parameters of the two spin states. This is well understood by writing the two order parameters in the form
$\Psi_\uparrow= \sqrt{n_\uparrow} \exp(i \phi_\uparrow)$ and  $\Psi_\downarrow= \sqrt{n_\downarrow} \exp(i \phi_\downarrow)$. The expectation value $E_{\Omega}= \langle H_\Omega\rangle$ of the operator $H_\Omega=-(\hbar \Omega/2)\sigma_x$ then takes the form 
\begin{equation} 
E_\Omega= -\frac{\hbar \Omega}{2}\int d{\bf r}(\Psi^*_\uparrow\Psi_{\downarrow}+\Psi^*_\uparrow\Psi_{\downarrow})=- \hbar \Omega \int d{\bf r}\sqrt{n_\uparrow n_\downarrow}\cos \phi_r \; ,
\label{EOmega}
\end{equation}
which depends explicitly on the relative phase $\phi_r$.  The breaking of this symmetry, which in the absence of Rabi or Raman coupling would add  to the $U(1)$ symmetry associated with the total phase of the two order parameters,  has deep consequences on the dispersion of the spin excitations,  causing the appearence of  a  gap (see  Sect.\ref{sec:excitationsRabi} and Sect.\ref{sec:excitationsSOC}).
The  dependence of $E_\Omega$ on the relative phase  also implies that the relative number $N_\uparrow-N_\downarrow=\langle \sum_j \sigma_{zj}\rangle$ of atoms in the two spin states is not conserved and obeys the  equation:
\begin{equation}
\frac{d(N_\uparrow-N_\downarrow)}{dt}= \frac{1}{i\hbar}\langle [\sum_j\sigma_{zj},H_\Omega]\rangle= \Omega \int d{\bf r}\sqrt{n_\uparrow n_\downarrow} sin \phi_r \; .
\label{dNrdt}
\end{equation}
The ground state of the mixture corresponds to the condition of equal phases ($\phi_\uparrow = \phi_\downarrow$) and hence to a stationary value of  $N_\uparrow-N_\downarrow$. Out of  equilibrium the relative number of atoms can instead  exhibit time dependent oscillations \cite{Cornell1998},  corresponding to the so called  internal Josephson effect (see Sect.\ref{sec:josephson}). The new topology imposed by the  relative phase dependence of $E_\Omega$  has also important consequences
on the nature of the solitonic solutions as well as on the rotational properties of the system and in particular on the behavior of the  vortex lines (see Sect.\ref{sec:domainwall}).
 
Both the Rabi  and the spin orbit Hamiltonians (\ref{hrf}) and (\ref{hSOC}) are translational invariant and commute with the canonical momentum 
$p_x=-i\hbar \partial_x$. In the case of SOC coupling the translational invariance can be spontaneously broken, giving rise to a peculiar stripe phase, with characteristic supersolid features (see Sect.\ref{sec:phasesSOC}).  
Translational invariance is not however equivalent to Galilean invariance  which is explicitly violated by the spin-orbit Hamiltonian. This is best understood   calculating how the spin-orbit Hamiltonian (\ref{hSOC}) is transformed by the unitary Galilean transformation 
$G=\exp(imvx/\hbar)$ which provides a Galilean boost, corresponding to the displacement $mv$ of the wave function in momentum space, along the $x$-direction. Only the $x$ component of the kinetic energy term is modified by the Galilean transformation and takes the form $ 
  G^{-1}(px-\hbar k_0\sigma_z)^2G/2m =  (p_x-\hbar k_0\sigma_z+mv)^2/2m$ 
so that, in the new frame,  the spin-orbit Hamiltonian $h^\prime_{\textrm{SOC}} =G^{-1}h_{\textrm{SOC}}G$   is given by 
\begin{equation} 
h^\prime_{\textrm{SOC}}= h_{\textrm{SOC}}+\frac{m}{2}v^2 +mv (p_x-\hbar k_0\sigma_z) \; .
\label{G2}
\end{equation}
The operator $(px-\hbar k_0\sigma_z)$, which represents the physical momentum of the particle, is not a constant of motion, because of the presence of the Raman coupling $\Omega \sigma_x$ in the spin-orbit Hamiltonian. As a consequence, the two Hamitonians $h^\prime_{\textrm{SOC}}$ and $h_{\textrm{SOC}}$ are physically different, yielding a violation of Galilean invariance with important consequences on the superfluid properties of the system, as discussed  in Sect.\ref{sec:superfluiditySOC}.

\section{Equilibrium and non equilibrium properties of Rabi coupled gases}

In this Section we describe the phase diagram, the elementary excitations, some topogical configurations, as well as far-from-equilibrium properties of a spinor condensate in the presence of Rabi coupling. The system has been thoroughly studied theoretically at the mean-field level (see \cite{RecatiReviewEPJD} and reference therein), the first studies dating  back to the 90`s  \cite{Goldstein1997,Blakie1999}. The first experimental realisation was obtained in the group of Eric Cornell \cite{RabiCornell99,RabiCornell2000} with the aim of studying  superfluidity in the presence of   spinor configuations.     

\subsection{Ground state properties}
\label{sec:gsRabi}

\begin{figure}
\centering
\includegraphics[width=0.5\textwidth]{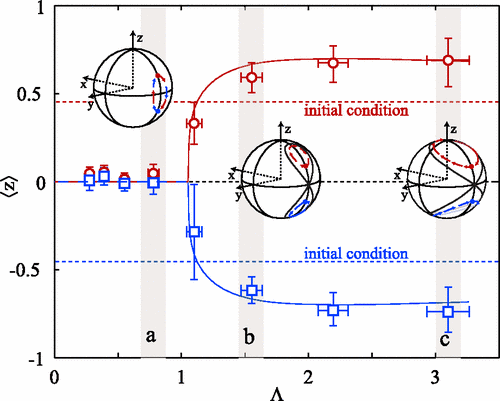}
    \caption{First experimental evidence of ferromagnetic-like bifurcation in a Rabi coupled gas of $^{87}$Rb atoms \cite{Zibold2010}. The total relative magnetisation $Z=(N_\uparrow-N_\downarrow)/(N_\uparrow+N_\downarrow)$ is plotted as a function of the ratio $\Lambda$, between the magnetic interaction energy and the Rabi coupling (see Sec. \ref{sec:josephson}). The Bloch spheres represent the dynamics around the fixed points of the Bose-Josephson junction Eqs. \ref{eq:BJJ}. }
    \label{fig:bifurcation}
\end{figure}

In the following we will mainly consider the case $\delta=0$, ensuring $\mathbb{Z}_2$ symmetry. In this case the mean field energy density of an homogeneous gas  reads
\begin{equation}
        \varepsilon_{MF}=\frac{g_{dd}}{2}n^2+\frac{g_{ss}}{2}s_z^2-\frac{\Omega}{2} \sqrt{n^2-s_z^2}\cos(\phi_r),
        \label{eq:energy}
\end{equation}
with $n=N/V$.  The stationary states are found by minimising the grand canonical energy $\varepsilon_{MF}-\mu n$ with respect to $n$,  $s_z$ and $\phi_r$, where  $\mu$ is the chemical potential. The ground  state is characterized  by the vanishing of the relative phase ($\phi_r=0$)  and obeys the coupled equations  
\begin{eqnarray}
\mu&=&g_{dd}n-\frac{\Omega n}{2\sqrt{n^2-s_z^2}}, \label{eq:mu}\\
0&=&g_{ss}s_z+\frac{\Omega s_z}{2\sqrt{n^2-s_z^2}}  \; ,
\label{eq:sz}\end{eqnarray}
exhibiting a bifurcation (see Fig. \ref{fig:bifurcation}) as a function of $\Omega$. While for $\Omega> -2g_{ss}n$ the ground state solution  has vanishing spin polarization, for smaller values of $\Omega$, requiring the condition $g_{ss} <0$,  the lowest energy solution corresponds to a typical ferromagnetic configuration with spin polarization 
\begin{eqnarray}
s_z=\pm n\sqrt{1-\left(\frac{\Omega}{2g_{ss}n}\right)^2} \; . 
\label{eq:ferro}
\end{eqnarray} 
The transition between the two regimes is   reminiscent of the quantum phase transition of  the Ising model in transverse field (see, e.g., \cite{Sachdev}) and for this reason the two states are referred to as paramagnetic    and  ferromagnetic  phases, respectively. The condition $\Omega+2g_{ss}n=0$, identifies the critical transition point. 
In the ferromagnetic phase the system will select one of the two polarisations, spontaneously  breaking the $\mathbb{Z}_2$ symmetry and the polarization, close to the critical point, grows 
 like $s_z\propto(-(2g_{ss}n+\Omega))^\beta$,  with the typical  mean field   critical exponent $\beta=1/2$.
 Also the  magnetic susceptibility $\chi=(\partial^2 \varepsilon_{MF}/\partial s_z^2)^{-1}$  exhibits the ferromagnetic behavior, diverging  near the critical point at  $\Omega+2g_{ss}n=0$.
 In the paramagnetic (P) and ferromagnetic (F)  phases one finds
\begin{eqnarray}
 \chi_P&=& \frac{2n}{2g_{ss}n+\hbar\Omega} \label{eq:chipara}\\
 \chi_{FM}&=&\frac{(\hbar\Omega)^2}{|g_{ss}|}\frac{1}{(2g_{ss}n)^2-(\hbar\Omega)^2}\label{eq:chiferro} \; 
\end{eqnarray}
respectively. 
In the absence of Rabi coupling the situation is very   different since the total polarisation $S_z=N_\uparrow-N_\downarrow$ is a conserved quantity, leading to a further $U(1)$ symmetry of the  Hamiltonian. In this case the two  ground states correspond  to the miscible phase if $g_{ss}>0$ (all the atoms occupy the same volume) and exhibit immiscibility if  $g_{ss}<0$, in which case  the two atomic species occupy distinct regions in space. At zero temperature an abrupt transition occurs as soon as the coupling constant $g_{ss}$ becomes negative. 

The first experimental measurement of the transition between a para- and a ferromagnetic phase in Rabi coupled gases 
was reported by the group of Markus Oberthaler \cite{Zibold2010,Nicklas2015} (see, Fig. \ref{fig:bifurcation}).  

\subsection{Elementary excitations of a Rabi coupled Bose-Einstein Condensed Mixture}
\label{sec:excitationsRabi}

Once the ground state is known, the excitations of the system are  determined by Bogolyubov theory. The Bogolyubov approach is  known to be equivalent to the solution of the linearized equations of time dependent Gross-Pitaevskii theory (see, e.g., \cite{SandroLevBook}).
In a quantum mixture the  equations for the elementary excitations  are more involved than in the single component case and for a more complete discussion we refer the interested reader to, e.g., \cite{Tommasini2003,RecatiReviewEPJD}. 
\begin{figure}
\includegraphics[width=1\textwidth]{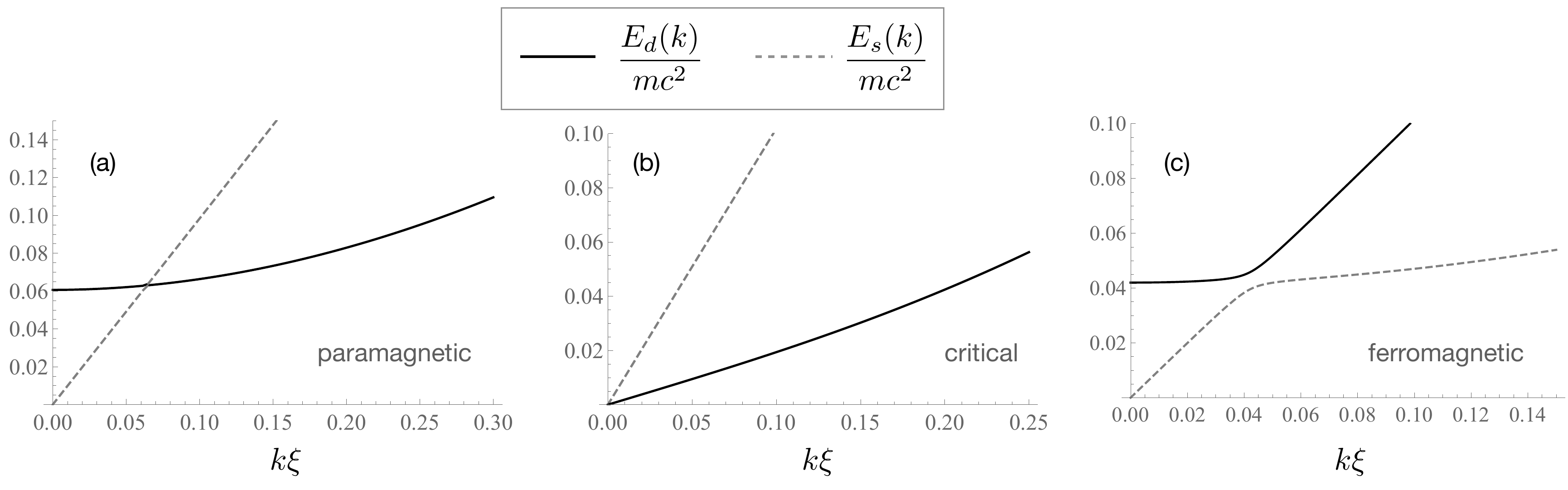}
    \caption{Spectrum of a Rabi coupled gas in the (a) paramagnetic and (c) ferromagnetic phase, and at the (b) critical point $\Omega+2g_{ss}n=0$.   
    }
    \label{fig:spectrumRF}
\end{figure}

In Fig.\ref{fig:spectrumRF} we report the typical form of the spectrum across the  ferromagnetic transition. The spectrum has two branches,  which  are usually referred to as density ($E_d(k)$) and spin  ($E_s(k)$) branches. The branch $E_d(k)$ is gapless 
and its low energy behaviour is dictated by the existence of the Goldstone mode due to the spontaneous breaking of the $U(1)$ symmetry related to the conservation of the total number of particles. At low momenta it 
has a phonon-like behaviour $E_d=c_d k$, with $c_d$  the  speed of density sound. 
The branch $E_s(k)$ is instead gapped, as a consequence of  the cost associated with the change of  the relative phase $\phi_r$. This cost  is enforced by the Rabi coupling which explicitly breaks the  symmetry $U(1)$ relative to the conservation of the relative atomic population, differently from what happens in  standard Bose-Bose mixtures (i.e. in the absence of Rabi coupling), where  also  the spin branch is gapless \cite{PethickBook}. 
According to the  theory of second order phase transitions the gap closes at the critical point and   has a different behaviour in the two phases  \cite{Sachdev}. We find:  
\begin{eqnarray}
     \Delta_P=\sqrt{2n\hbar\Omega\chi_P^{-1}}\rightarrow\left(\hbar\Omega(2g_{ss}n+\hbar\Omega)\right)^{1/2}\;{\textrm{ for}}\;2g_{ss}n+ \hbar\Omega\rightarrow0^+ \label{eq:gappara}\\
 \Delta_{FM}=\sqrt{\frac{2(\hbar\Omega)^2}{|g_{ss}|}\chi_{FM}^{-1}}\rightarrow\left((2\hbar \Omega|2g_{ss}n+\hbar\Omega|)\right)^{1/2}\;{\textrm{ for}}\;2g_{ss}n+\hbar\Omega\rightarrow0^-.\label{eq:gapferro}
\end{eqnarray}
where we can identify  the   critical exponent for the gap which,  within the present mean-field theory, coincides with the exponent $\beta=1/2$ for the magnetisation close to the critical point (see Eqs. \ref{eq:ferro}).   
At the critical point the dispersion relation becomes linear with the speed of spin sound given by $c_s=\sqrt{|g_{ss}|n/m}$. Notice that the spin spectrum for a standard Bose-Bose mixture becomes instead quadratic at the transition point to the immiscible regime, while  in the   phase separated regime  the concept of   spin sound  does not make sense anymore.

While the spin spectrum is gapless both for $\Omega=0$ and at the critical point $\hbar\Omega=-2g_{ss}n$,  the quantum fluctuations associated with the corresponding long wavelength modes behave very differently in the two cases. For $\Omega=0$ the linear low-$k$ energy mode is dominated by the fluctuations of the \textit{relative phase}. On the other hand at the critical point the low energy mode is dominated by the fluctuations of the \textit{relative population (polarisation)},   reflecting the critical nature  of the ferromagnetic transition as we will now discuss. 

In the unbroken $\mathbb{Z}_2$ paramagnetic  phase, the Bogolyoubov predictions for the dispersion law and for the corresponding quantum fluctuations take a  particularly simple and instructive form.. 
In this phase  the   fluctuations of the total density ($\Pi_d$) and spin density ($\Pi_s$) operators as well as of the  total phase ($\phi_d$) and relative phase ($\phi_r$) operators  can be explicitly written in terms of the annihilation (creation) operators $d_\vk$ ($d^\dagger_\vk$)    and  $s_\vk$ ($s^\dagger_\vk$) for the density and spin excitations respectively, as:



\begin{eqnarray}
\Pi_\alpha(\vr)&=&\frac{\sqrt{n}}{2}\sum_\vk ({ U}_{\alpha,k}+{ V}_{\alpha,k})(\alpha_\vk e^{i\vk\cdot\vr}+\alpha^\dagger_\vk e^{-i\vk\cdot\vr}), 
\label{Pi}
\\
\phi_\alpha(\vr)&=&i\sqrt{1\over 2n}\sum_\vk ({ U}_{\alpha,k}+{ V}_{\alpha,k})^{-1}(\alpha_\vk e^{i\vk\cdot\vr}-\alpha^\dagger_\vk e^{-i\vk\cdot \vr}),
\label{fi}
\end{eqnarray}
where $\alpha=d,\,s$ and $\vk$ is the momentum of the corresponding excitations.  In the above equations we have introduced the so called Bogolyubov amplitudes $U$'s and $V$'s, whose combination provides the contribution of each mode to the static structure factor according to $S_{\alpha}\propto |U_{\alpha,k}+V_{\alpha,k}|^2$.
The proper diagonalization of the Bogoliubov Hamiltonian  yields the following results for the Bogoliubov amplitudes of the density and spin excitations \footnote{see Ref.~\cite{Tommasini2003} for the most general case  $g_{sd}\ne 0$. }.
\begin{equation}
U_{d,k}+V_{d,k}=\left({k^2\over k^2+8mg_dn}\right)^{{1\over 4}},
U_{s,k}+V_{s,k}=\left({k^2+2m\hbar\Omega \over k^2+4m(2g_{ss}n+\hbar\Omega)}\right)^{{1\over 4}}.
\label{Us}
\end{equation}
Analogously, one finds the following expressions for the dispersion laws: 

\begin{eqnarray}
 \label{2orderDiag}
 E_d(k)&=&\sqrt{\frac{\hbar^2k^2}{2m}\left(\frac{\hbar^2k^2}{2m}+2g_{dd}n\right)},\\
 E_s(k)&=&\sqrt{\left({\hbar^2k^2\over 2m}+\hbar\Omega\right)\!\!\left({\hbar^2k^2\over 2m}+2g_{ss}n+\hbar\Omega\right)}\;,
\end{eqnarray}
allowing, in the $k\to 0$ limit, for the identification  of the density sound velocity    $c_d=\sqrt{g_{dd}n/m}$  and of the spin gap $\sqrt{ \hbar \Omega ( 2g_{ss}n+ \hbar \Omega)}$.

When $\Omega=0$ and $g_{ss}>0$  the Bogoliubov amplitudes in both the density and spin channels have the same structure as for the single component Bose gas and, as $k \to 0$,   the fluctuations of the phase  diverge, while the static structure factors  vanish linearly  due to atom number conservation in each component.

If $\Omega >0$  the spin channel  instead reveals  a very different behaviour. The phase and amplitude mode are generally comparable also in the long-wave length limit, corresponding to a finite value of the spin static structure factor at low momenta. More importantly,  at the critical point the fluctuations of the spin density  become critically large,  providing the ${k^{-1}}$ divergent behavior of the spin structure factor, consistently with the divergent behavior of the magnetic polarizability Eq.(\ref{eq:chipara}) (see also Table 1).  Such a critical behaviour of the spin fluctuations has been predicted to lead to a strong damping of the (density) Goldstone phonons. Indeed, while the Bogoulyubov  approach predicts an infinite life-time for the elementary excitations, the phonon modes can decay into two lower energy phonons leading to the so called Belyaev damping which scales as $\Gamma_{ddd}(k)\rightarrow k^5$ at small momenta. On the other hand the closing of the gap opens a new decay channel, where a density mode can decay into two spin modes, yielding an enhanced damping which scales as $\Gamma_{dss}(k)\rightarrow k$ \cite{RecatiPiazza2019}.


\begin{table}[h!]
\centering
\begin{tabular}{|p{2cm}||c|c|c|}
 \hline
  & Rabi & Spin-Orbit & Incoh. Mixture\\
 \hline
 $\omega_{s,k}$ & $\sqrt{ |\Omega-\Omega_c|}$ & $\sqrt{|\Omega-\Omega_c|}k$  &  $c_s k$ \\
 $| \langle 0|\hat{s}_{z}(\mathbf{k})|n\rangle|^2$  & $\Omega|\Omega-\Omega_c|^{-1/2}$    &$k|\Omega-\Omega_c|^{-1/2} $ & $k/c_s$\\
 $M_{-1,s}$ & $|\Omega-\Omega_c|^{-1}$ & $|\Omega-\Omega_c|^{-1}$  & $c_s^{-2}$ \\
 
 \hline
\end{tabular} 
\vspace{0.4cm}
\caption{\color{black} First row: dispersion of the low frequency modes excited
by the spin operator $s_z(k)$ at small $k$,
near the Paramagnetic-Ferromagnetic phase
transition at $ \Omega =\Omega_c$, for both the
Rabi and the Spin-Orbit coupled mixtures. The
spin strength $|\langle 0|s_z(k)|n\rangle|^2$ 
as well as the
corresponding contributions to the
magnetic susceptibility sum rule $M_{-1}$ are
reported, in the 2nd and 3rd row, respecively. For the sake of completeness, in the third
column we report the results for incoherent
mixtures, where $c_s= \sqrt{g_{ss}n/m}$. In the
Rabi case the spin dispersion is gapped, unless
one works exactly at the transition (see Sect.
3.2). In the spin-orbit case the dispersion of
the low frequency mode is instead linear in $k$
and the sound velocity becomes softer and
softer as $\Omega \to \Omega_c$ (see Sect.4.3).
Remarkably, the two models instead predict
the same divergent behavior of the magnetic
susceptibility at the transition.}
\label{table}
\end{table}

 \subsection{Hydrodynamic formulation and internal Josephson effect}\label{sec:josephson}

In the  previous Sections we have discussed the  ground state properties of   Rabi coupled gases and the  small  amplitude oscillations around equilibrium.
We provide a more general description of the  mean-field dynamics of the spinor gas, by developing the hydrodynamic formulation of the Gross-Pitaevskii equations. This   formulation emphasizes in a explicit way the role of the spin density. 
In $s=1/2$ spinors the spin density components are defined by 
$s_i(\mathbf{r})= (\Psi_\uparrow^*,\Psi_\downarrow^*)\sigma_i(\Psi_\uparrow,\Psi_\downarrow)^T$, with $\sigma_i$, $i=x,y,z$ the Pauli matrices.  In  particular $s_x=\sqrt{n^2-s_z^2}\cos\phi_r$,  $s_y=\sqrt{n^2-s_z^2}\sin\phi_r$ and the relation $|\mathbf{s}({\bf r})|=n({\bf r})$ holds.  The velocity field,  defined as the total current divided by the density, takes the simple form 
\begin{equation}\V{v}(\mathbf{r})=\frac{\V{j(\mathbf{r})}}{n}=\frac{\hbar}{2m n i}\sum_{\sigma=\uparrow,\downarrow}
(\Psi^*_\sigma\nabla\Psi_\sigma-\Psi_\sigma\nabla\Psi_\sigma^*)=\frac{\hbar}{2m}(\nabla \phi_d+s_z/n\,\nabla\phi_r),
\end{equation}
where $\phi_{d(r)}=\phi_\uparrow\pm\phi_\downarrow$ is the total (relative) phase. Due to  the spinor nature of the wave function  the velocity field $\V{v}(\mathbf{r})$ is not  in general irrotational, but satisfies the relation  $\nabla\times v=\hbar/(2m)\nabla (s_z/n)\times \nabla \phi_r$, corresponding to the  analogous of the Mermin-Ho relation \cite{MerminHo75} originally introduced for describing the superfluid A-phase of $^3$He.   Eventually the hydrodynamics equations can be written as (see, e.g., \cite{Nikuni2003}): 
\begin{eqnarray}
 & \dot{n}+\D(n \V{v})=0,  \label{eq:shydro-cont}\\
 &m\dot{\V{v}}+\nabla\left( \frac{mv^2}{2}+\mu+\frac{s_z}{n}h+V-\frac{\hbar^2\nabla^2\sqrt{n}}{2m\sqrt{n}}+\frac{\hbar^2|\nabla\mathbf{s}|^2}{8m n^2}\right)=0,\label{eq:shydro-euler}\\
&\dot{\mathbf{s}}+\sum_{\alpha=x,y,z}\partial_\alpha(\V{j}_{s,\alpha})=\mathbf{H}(\V{s})\times{\V{s}},\label{eq:spin}
\end{eqnarray}
where, for completeness,  we have included a possible external trapping potential $V$. 

The first equation is the standard continuity equation for the particle number conservation and  the second one is the Euler equation, with  the chemical potential $\mu$ and the internal magnetic field $h$ given by 
\begin{equation}
\label{muandh}
\begin{split}
    \mu&=g_{dd}n-\frac{\hbar\Omega}{2}\frac{n}{n^2-s_z^2}s_x\, ,\\
    h&=g_{ss}s_z+\frac{\hbar\Omega}{2}
    \frac{s_z}{n^2-s_z^2}s_x\;.
\end{split}
\end{equation}
Notice that there is no problem with the limiting case $s_z\rightarrow \pm n$, i.e. a fully polarised mixture since the term $\mu+ hs_z/n =g_{dd}n+g_{ss}s_z+\hbar\Omega s_x/(2n)$ entering in the second Euler equation is well defined for any $s_z$.  
In the lower energy states, where  $s_x=\sqrt{n^2-s_z^2}$, and $h=0$ the above equations reduce to \aref{eq:mu} and \aref{eq:sz} and the corresponding   susceptibilities are given by  $\chi^{-1}=\partial h/\partial s_z$. As pointed out in \cite{Nikuni2003}, despite the possible presence of rotational components in the velocity field $\V{v}$, the time derivative $\dot{\V{v}}$ turns out to be irrotational. 

The last \aref{eq:spin} is the most interesting one, since it determines the spin dynamics. 
The l.h.s. is the continuity equation, due to the  Noether theorem for the $SU(2)$ symmetry. The spin current  contains two contributions:
\begin{equation}
    \V{j}_{s,\alpha}=v_\alpha\mathbf{s}-\frac{\hbar}{2m}\left(\frac{\mathbf{s}}{n}\times\partial_\alpha \mathbf{s}\right), \; \;  \alpha=x,y,z\;,
\end{equation}
the first term being  the classical spin advection and the second one corresponding  to the spin-twist, whose contribution to the equation of motion is called \textit{quatum torque}.
In our system the $SU(2)$ symmetry is reduced to the  $\mathbb{Z}_2$ symmetry by the effective field $\mathbf{H}(\V{s})=(-\Omega,0,2 g_{ss} s_z/\hbar)$ which enters  the r.h.s. of \aref{eq:spin}.

If the dynamics involves neither the density nor the  velocity, the system is described only by the spin \aref{eq:spin}. In this case the equation of motion for a coherently coupled BEC is formally equivalent to a dissipationless version of the so-called Landau-Lifshitz equation (LLE) for the magnetisation dynamics in ferromagnets (see, e.g., \cite{Baryakhtar2015} and reference therein). For  uniform  configurations, where the divergence of the spin-current is negligible, the equations take the form 
 $\dot{\V{s}}=(-\Omega,0,2g_{ss}s_z/\hbar)\times\V{s}$, 
also called Bose Josephson Junction (BJJ) equations \cite{Smerzi1997}. 
 In terms of the relative magnetization $Z=s_z/n$ and of the relative phase $\phi_r$ they can be written in  the form:
\begin{eqnarray}
    \dot{Z}&=&-\Omega\sqrt{1-Z^2}\sin\phi_r\\
    \dot{\phi_r}&=&\Omega Z\left(\Lambda+\frac{1}{\sqrt{1-Z^2}}\cos\phi_r\right) \; 
    \label{eq:BJJ}
\end{eqnarray}
where $\Lambda=\frac{2g_{ss}n}{\hbar\Omega}$. Depending on the initial condition and on the value of $\Lambda$,   the BJJ equations
exhibit a number of different dynamical regimes (we refer the reader to original theory Refs. \cite{Smerzi1997,Raghavan1999} for a more comprehensive discussion). In the present context it is useful to remind that the  stationary points are characterised by $\phi_0=0\;\mathrm{or}\;\pi$ and by the value  $Z_0$ of the spin polarization. 
The value of $Z_0 $ is different from $0$ only for $|\Lambda|>1$, with $Z_0=\pm\sqrt{1-\Lambda^{-2}}$.
In the limit of small amplitude oscillations around  equilibrium  the system is characterized by the frequency:
\begin{equation}
    \omega_{J}^2= \Omega^2 \left(\Lambda\sqrt{1-Z_0^2}\cos{\phi_0} + \frac{1}{1-Z_0^2}\right),
    \label{eq:omegaJJ}
\end{equation}
which, in the absence of interactions ($\Lambda =0$) reduces to $\omega_{J}= \Omega$. 

The paramagnetic and ferromagnetic phases described in the previous Sections correspond to the stationary solutions  $\phi_0=0, Z_0=0$ and   $\phi_0=0, Z_0=\pm\sqrt{1-\Lambda^{-2}}$ , respectively. In both cases the value $\hbar\omega_J$ coincides with the gaps of the spin elementary excitation for  $k\to 0$.   The stationary point    $\phi_0=\pi$ instead corresponds to a maximum of the  Rabi energy \ref{EOmega} and gives rise to a dynamically unstable configuration, the excitation spectrum,  derivable from the linear solutions \aref{eq:shydro-cont}-\aref{eq:spin},  becoming imaginary for some values of the momentum.

The existence of stationary points with $Z_0\neq 0$ is at the origin of the celebrated self-trapping regime  \cite{Smerzi1997,Raghavan1999}. In such a regime the polarisation cannot change sign, i.e., if the majority of atoms is initially in one of the two spin states state, this will be true during all the time evolution. Self-trapping configuration in which both the phase and the polarisation oscillates are called $0$ or $\pi$-modes, depending on the value of $\phi_0$. More interesting are the so-called running phase modes, in which the phase keeps increasing. Running phase solutions exist only for $|\Lambda|>2$ and in this case the phase space $Z-\phi$ is reminiscent of the angle-angular velocity phase space of a classical pendulum. In particular there exists a separatrix, between the (Josephson) oscillating regimes -- with zero time average polarisation -- and the self trapped regimes. The period of the oscillations diverges approaching the separatrix and the effect is also named critical slowing down.  

The variuos BJJ regimes have been experimentally  investigated in detail both in the originally proposed \cite{Smerzi1997} double-well potential geometry  \cite{Albiez2005,Schumm2005,Levy2007,Trenkwalder2016,Spagnolli2017} -- where the imbalance $Z$ corresponds to the difference in the number of atoms in the right and in the left well -- as well as in the internal Josephson configuration \cite{Zibold2010,Nicklas2011}, using Rabi coupled BEC's.

 It is worth mentioning that the dynamic instability of the  $\phi_0=\pi$ configuration is not relevant for most of the present day experimental realisations of the BJJ equations. Indeed the atoms are usually trapped by tight confinements, so that the orbital degrees of freedom are frozen out and the BJJ equations properly describe the dynamics of the system \footnote{In the case of coherently coupled BEC's the only change in \aref{eq:BJJ} is 
 $Z=(N_\uparrow-N_\downarrow)/N$ and $\Lambda=2g_{ss}N\alpha/(\hbar\Omega)$, with $N_\sigma$ the total atom in the state $\sigma$, $N=N_\uparrow+N_\downarrow$ and $\alpha$ a constant which takes into account the shape of the wave function of the tightly trapped gas (see, e.g., \cite{Zibold2010})}.

  
The instabilities occurring  in extended systems, where one should rather use the full Eq. \ref{eq:shydro-cont}-\ref{eq:spin} are of the  Cross-Hohenberg type \cite{Cross1993} and  are characterised by complex elementary excitations at rather well defined momenta and whose energies have a vanishing or finite real part, and for this reason are called $I_s$ (static) and $I_o$ (oscillatory) instabilities, respectively \cite{Cross1993}.   

In cold gases mixtures pattern formation due to an $I_s$-type instability has been already observed in spin-1 mixtures by Sengstock's group \cite{MatusIs,SengstockIs}.
A proposal to observe the instability $I_o$ in coherently coupled gas has been put forward in \cite{DallaTorreInst} by means of a sudden quench from the stable ($\phi_r=0$) to the unstable  ($\phi_r=\pi$) configuration.   

Let us conclude this Section by mentioning that, while Rabi coupled gases have been used to simulate the BJJ equations, at the moment the simulation of the full LLE Eq. (\ref{eq:spin}), where one takes into account both the time and the position dependence of the spin density, has not been explored.
One of the main reason is the experimental difficulty in having a low and stable Rabi coupling in large   systems, in order to reveal the interplay between the $x$ and $z$ components of the effective magnetic field $\V{H}$ together with the position dependence of the spin density. However very recently, the possibility of describing Rabi coupled gases with the LLE equations has been experimentally verified and used to study the effect of the critical slowing down of the Josephson dynamics at the separatrix between a region in the self-trapped regime and a region in the Josephson oscillation regime \cite{TrentoQtorque}.

\subsection{Topological excitations: relative phase domain walls and half-quantum-vortices}
\label{sec:domainwall}

Further peculiar features exhibited by Rabi coupled mixtures concern  the phenomena related to  topological defects, like solitons and vortices, involving the relative phase of the two components.   A remarkable example  is the relative phase domain wall,  originally  identified  by Son and Stephanov \cite{SonDVortex}  \footnote{The same kind of solution has been also described in the context of two-band superconducors by Tanaka in  \cite{Tanaka}}. As we will discuss later the  existence of this soliton solution is  deeply  connected with the novel features exhibited by quantized vortices in the  mixture.
 A simple description can be obtained in the paramagnetic phase of  uniform matter,  under the assumption $\hbar\Omega \ll g_{ss}n\ll  g_{dd}n$ \cite{Tylutki2017}.  In this limit both the density $n_\uparrow$ and $n_\downarrow$ can be regarded as uniform and equal, the only important degrees of freedom being the phases of the two order parameters, or better the total $\phi_d=\phi_\uparrow+\phi_\downarrow$ and the relative phase $\phi_r=\phi_\uparrow-\phi_\downarrow$.
The energy of the system, apart from a constant term,   then takes the form
\begin{equation}\label{eq:Ephiupdown}
E(\phi_d,\phi_r)= \frac{n}{2}\int d{\bf r}  \left[\frac{\hbar^2}{4m}\left(\nabla\phi_d\right)^2+ \frac{\hbar^2}{4m} \left(\nabla\phi_r\right)^2 -\hbar \Omega\cos(\phi_r)\right] \; ,
\end{equation}
yielding the differential sine-Gordon equation  
${\hbar}\nabla^2\phi_r=2m\Omega\sin\phi_r$ \cite{SonDVortex,Tanaka}. For the domain wall solution, $\phi_r$ is a function of only one coordinate, say $x$, with the boundary condition that $\phi_r$ approaches a constant value as $x\to \pm \infty$. The trivial solution 
$\phi_\uparrow=\phi_\downarrow +2\pi m= const$,  with $m$ integer, corresponds to the ground state solution of the Gross-Pitaeveskii equations.  A non trivial solution, corresponding to an infinite domain wall located at $x=0$, is given by (see, e.g., Ref.~\cite{Tabor1989})
\begin{equation}
\phi_{r}(x)=4 \arctan\left(\exp(x/\xi_\Omega\right)\;,
\label{eq:eq2}
\end{equation}
 whose spatial variation is characterized by the Rabi healing $\xi_\Omega =\sqrt{\hbar/m\Omega}$. This stationary solution -- which is a local minimum of the energy -- connects two asymptotic ground states as  $x$ goes from $- \infty$ yo $+\infty$ with relative phase equal to $0$ and  $2\pi$ in the two limits,  respectively.

The  solution Eq.(\ref{eq:eq2}) generates a counterflow current and accumulates the relative phase gradient in a small region of size $\xi_\Omega$. The tension of the domain wall, i.e. its energy per unit area,  is equal  to $\sigma=(2\hbar)^{3/2}\sqrt{\Omega/m}$, revealing that the creation of a relative phase domain wall of infinite lengh would cost an infinite energy amount. 

\begin{figure}[h!]
\centering
\includegraphics[width=1\textwidth]{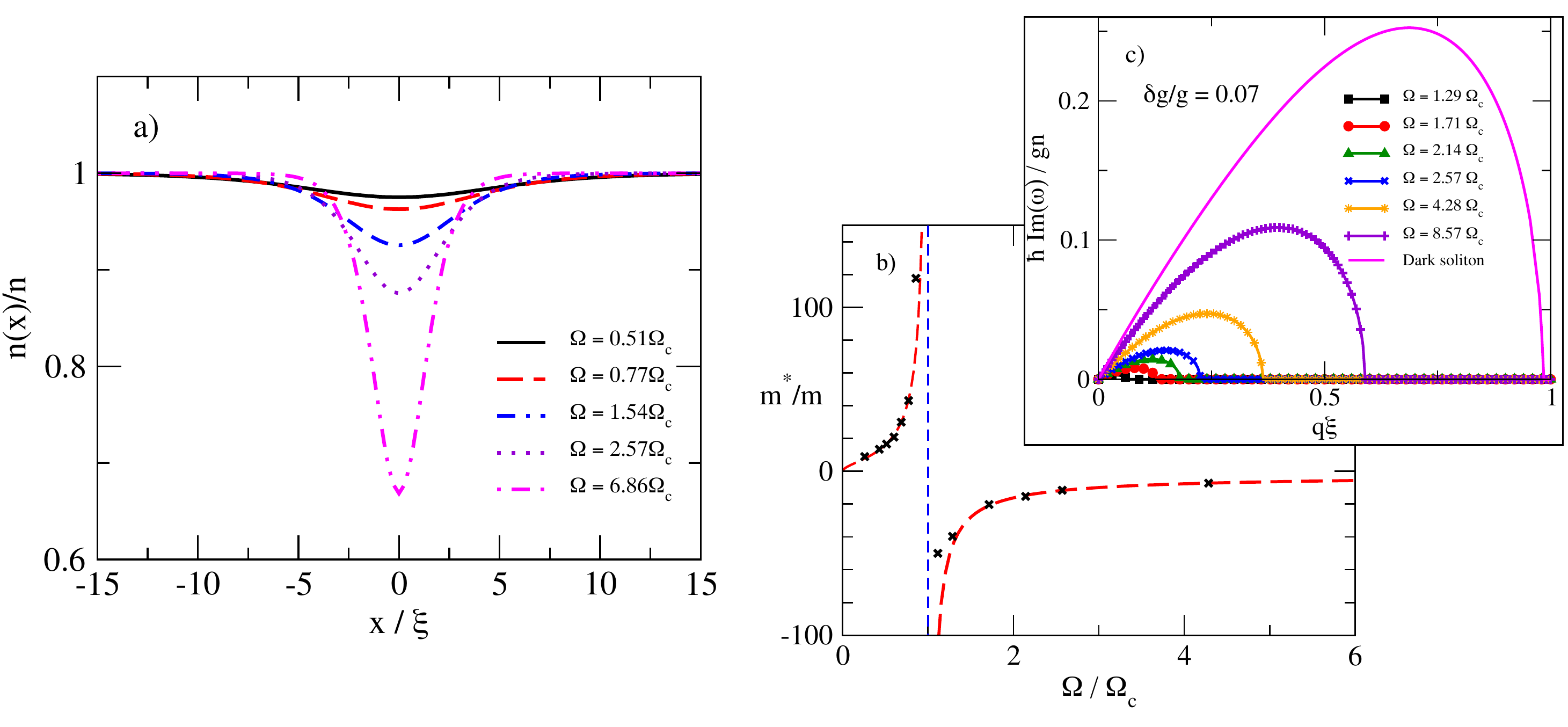}
    \caption{Relative phase domain wall obtained by solving coupled the Gross-Pitaevskii equation as a function of the coupling $\Omega$: a)  density profile; b) effective mass and c)  non-zero imaginary part of the spectrum yielding dynamic instability for $\Omega>\Omega_c$. From \cite{GallemiDecay}.}
    \label{fig:dwprofile}
\end{figure}

As already mentioned Eq.(\ref{eq:eq2})  is an approximation. However the solution of the more general Gross-Pitaevskii equations for the stationary solitonic solution exhibits a similar  equivalent phase pattern, but with a density dip near the wall, as a consequence 
of the compressible nature of the gas. The dip increases with the increase of the Rabi 
coupling~\cite{Usui2015}, as  shown in Fig. \ref{fig:dwprofile} a), and for large values of $\Omega$ 
the central density vanishes, in close analogy with dark solitons.

An important question to discuss concerns the stability of the domain wall, which exhibits a deep difference with respect to  the solitonic solution of a single component  Bose-Einstein condensate. In the latter case the soliton is well known to suffer dynamic  snake instability, unless one strictly works in 1D \cite{SandroLevBook}. As  pointed out by Son and Stephanov  \cite{SonDVortex}, for values of  $\Omega$ below a critical value $\Omega_c$, the domain wall solution is instead dynamically stable even in 3D configurations.  Under the assumption $\hbar \Omega, g_{ss}n \ll g_{dd}n$ the critical value takes the simple expression $\Omega_c= ng_{ss}/3$ \cite{SonDVortex}. The stability of the domain wall  is the consequence of the positiveness of its effective mass  \cite{GallemiDecay,KasamatsuDecay} whose dependence on the Rabi coupling is shown in  Fig. \ref{fig:dwprofile} b).   Close to $\Omega_c$ the effective mass exhibits a divergent behavior, while  for larger values  it  becomes negative and the domain wall undergoes dynamical snake instability, as a consequence of the appearence of an imaginary part in the excitation spectrum (see  Fig. \ref{fig:dwprofile} c). 

Let us now discuss the consequences of the relative phase domain wall on the structure of  vortex lines and let us consider a hypothetical domain wall of finite length.
Around its end points the relative phase must change by $2\pi$, corresponding to  the presence, at the end point of the domain wall,  of a vortex line in one of the two components, usually called half-quantum vortex (HQV). This reveals that    half-quantum-vortices  (HQV) cannot exist as  isolated objects \cite{UedaDVortex}, but they are always linked to a domain wall,  eventually ending with a second HQV of the same atomic  species with  opposite circulation, or with a HQV of  the other species with the same circulation, thereby forming a sort of vortex molecule.  This situation -- as already pointed out in \cite{SonDVortex} -- has  intriguing  analogies with  the quark confinement  in the theory of strong interactions \cite{Nitta2018}.
 
	\begin{figure}
\centering
\includegraphics[width=0.8\textwidth]{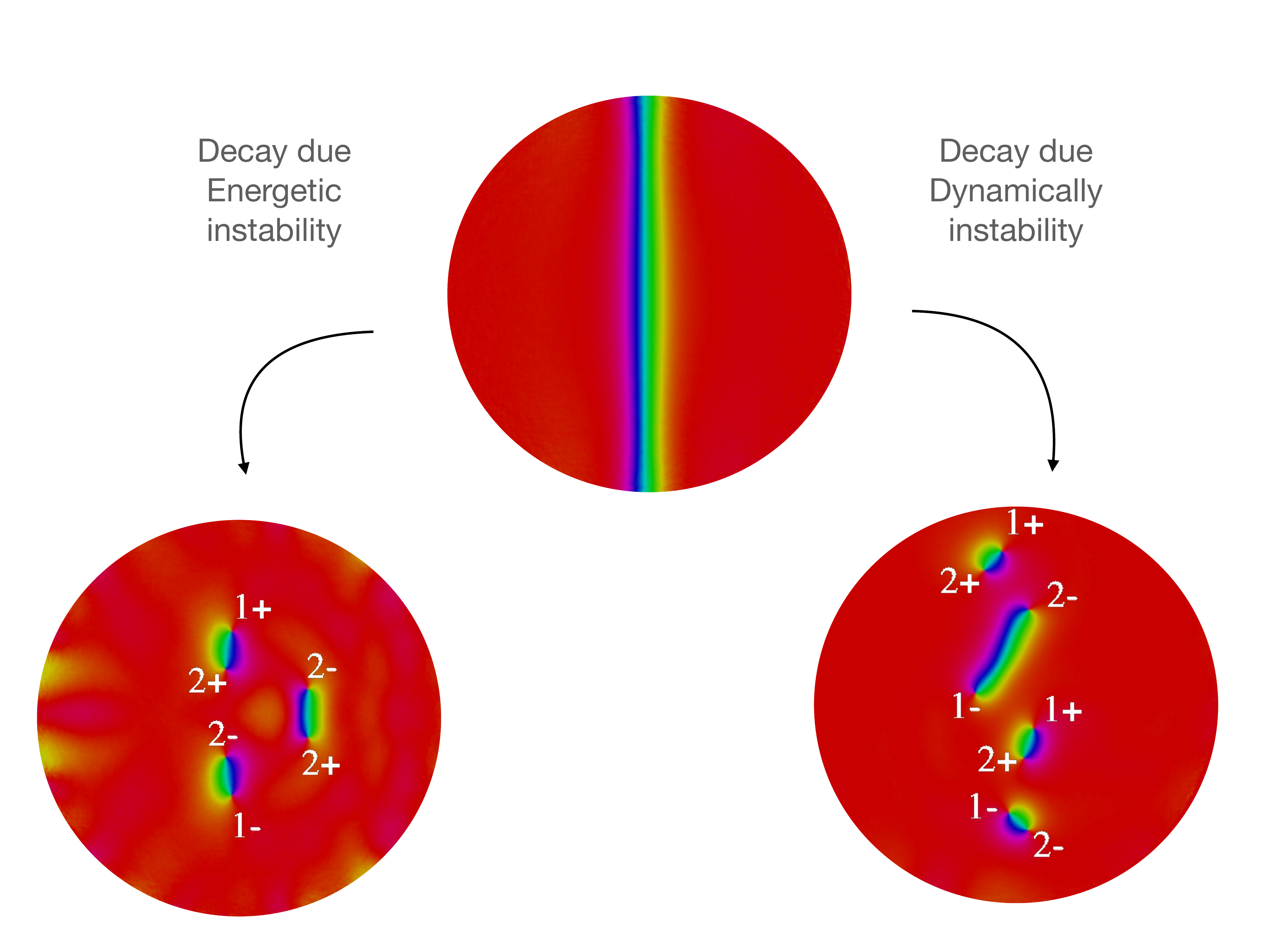}
    \caption{Phase profile of a long domain wall created at $t=0$ (upper panel) across a two-dimensional Rabi coupled gas trapped by an harmonic potential $V(x,y)=1/2 m \omega^2_{\textrm{ho}}(x^2+y^2)$. The circle contour correspond to the radius of the cloud where the Thomas-Fermi density is zero. The domain wall decays (left) via energetic instability, after bending,  into three composite objects or (right) via snake dynamic instability into four composite objects.  The pairs are either made of vortices with the same circulation ($\pm$ in the figure) in  two different components $1$ and $2$ or with opposite circulation in the same component. For more details see  \cite{Nitta2018,GallemiDecay}}
    \label{fig:dwdecay}
\end{figure} 
 
It is remarkable that such a classical field theory presented here can give rise not only to the confinement of HQV’s but also to the pair creation phenomenology typical of  Quantum Chromodynamics. A HQV-pair is indeed stable only if its size is smaller then a critical value   \cite{Nitta2018}. If  the pair   size is larger than this critical  value, it decays into two (or more) composite objects \cite{Tylutki2016,Nitta2018,GallemiDecay}. The actual decay mechanism depends on a number of parameters, in particular on the value of the Rabi coupling  $\Omega$. In Fig. \ref{fig:dwdecay} we report the case of a domain wall created across a cloud trapped by an harmonic potential where the density becomes smaller and smaller as one approaches the surface region. The figure reports the case of both a dynamically stable ($\Omega < \Omega_c$) and a dynamically unstable ($\Omega >\Omega_c$) configuration. In the former case the decay mechanism is very slow, while in the latter one the snake instability proceeds in a very fast way. In both cases the formation of the vortex pairs, after fragmentation of the domain wall,  preserves the initial vanishing value of  angular momentum \cite{GallemiDecay}.

To our knowledge there exists at the moment only a single experiment \cite{ShinComposite} reporting  the observation a vortex-domain wall  composite object in cold atomic gases. Although it has been obtained in a spin-1 mixtures, the observed configuration exhibits important analogies with the scenario described above for a spin-1/2 Rabi mixture.

\section{Spin-orbit coupled  configurations}

In this Section we describe the quantum phases, the elementary excitations and the superfluid properties of spin-orbit coupled Bose-Einstein condensates, emphasizing   analogies and   differences with respect to the Rabi coupled mixtures discussed in the previous section.  Spin-orbit coupled BEC's   were first experimentally investigated in the pioneering papers by the Spielman team 
\cite{Lin2009,Lin2011}, motivated by the possibility of generating synthetic gauge fields (see also \cite{SpielmanReview,DalibardGaugeFields,Galitski2019}). Experimental results for degenerate spin-orbit Fermi gases have been also soon become available \cite{SOCFermi,ZwierleinFermions}.  Theoretically, the study of SOC gases has been the object of extensive investigations in the last years. For previous review papers see \cite{zhaiReview, Li2015review}. From the many-body point of view particularly challenging features are exhibited by the so called stripe phase, where first important supersolid effects have been already identified experimentally \cite{Li2017,Putra2020}. 
	
\subsection{Single particle excitation spectrum}
\label{spSOC}
	
A discussion of spin-orbit coupling in BEC systems naturally starts  from the study  of the excitation spectrum of the single particle Hamiltonian (\ref{hSOC}). The deep modifications induced by   spin orbit coupling on the single-particle properties  are in fact crucial to understand the novel many-body features.   
	
\begin{figure}
\centering
\includegraphics[width=0.9\textwidth]{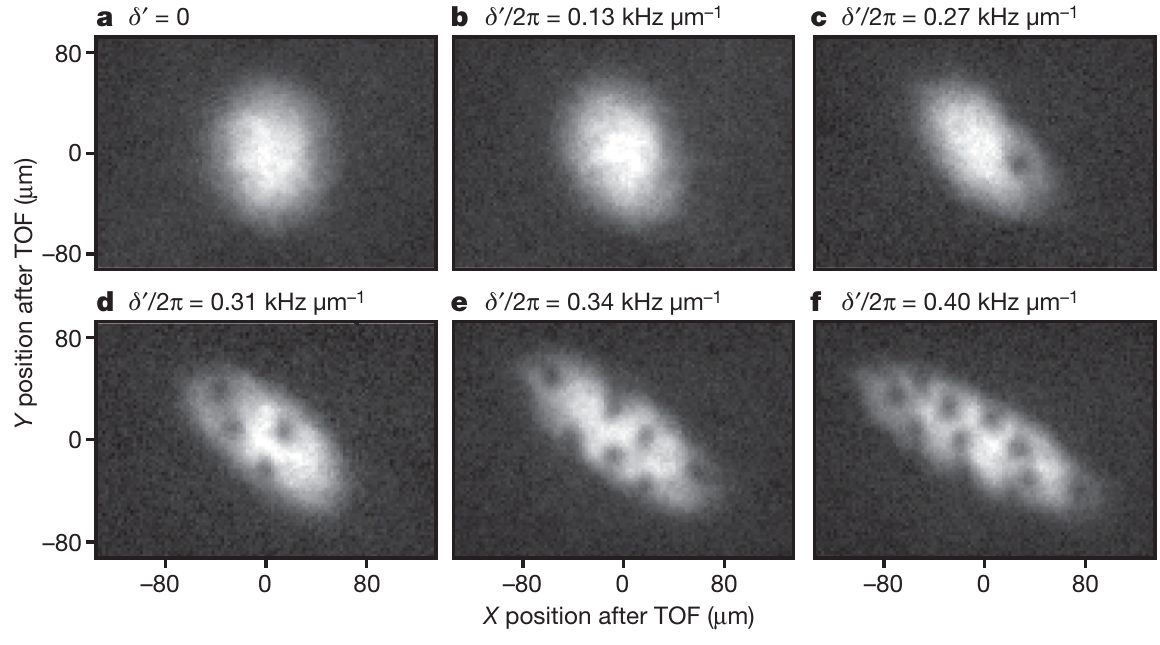}
    \caption{Appearance of vortices in a SOC trapped  BEC   containing 
N = 1.4 × 10*5 Rb atoms, at different detuning gradients. From \cite{Lin2009} }
    \label{vorticesSpielman}
\end{figure}

In uniform matter, where the canonical momentum ${\bf p}$ is a good quantum number, the  eigenvalues of Eq.(\ref{hSOC}) are given by:
\begin{equation}
\epsilon_{\pm}({\bf p})= \frac{p^2_x+p^2_\perp}{2m}+ E_r\pm \hbar \sqrt{\left(\frac{k_0p_x}{m}-\frac{\delta}{2}\right)^2+\frac{\Omega^2}{4}} \; ,
\label{epsilonssp} 
\end{equation}
where $E_r= (\hbar k_0)^2/2m$ is the recoil energy and $p^2_\perp \equiv p^2_y+ p^2_z$. The dispersion exhibits a typical double-band structure, reflecting the spinor nature of the configuration. Most interestingly, the dispersion  is characterized by the occurrence of a double minimum for small $\Omega$ and $\delta$, with the possibility of hosting BEC in single particle states with $p_x\ne 0$. For large values of $\Omega$ (much larger than the recoil energy $E_r$), the lower branch exhibits a single-minimum structure of the form
\begin{equation}
\epsilon({\bf p})\to \frac{ p^2_\perp}{2m}+ \frac{1}{2m}(p_x + \hbar k_0\delta/\Omega)^2 + const \; .
\label{epsilonssplargeOmega} 
\end{equation}
	By introducing a  space dependence in the   detuning $\delta$ of the form $\delta= \alpha y$, the new gauge field in (\ref{epsilonssplargeOmega}) is responsible for an effective uniform magnetic field $B=\alpha \hbar k_0/\Omega$,  oriented along the $z$-direction. This possibility was implemented experimentally in the pioneering work \cite{Lin2009} to generate an effective Lorentz force, responsible for the appearence of quantized vortices in BEC gases (see  Fig.\ref{vorticesSpielman}).

In the case of vanishing detuning $\delta$ the single particle dispersion (\ref{epsilonssp}) exhibits, for $\Omega < 4 E_r$, two symmetric minima at quasi-momentum $p_x=\pm\hbar k_1$ with $k_1=k_0\sqrt{1-(\hbar\Omega/4E_r)^2}$. For larger values of $\Omega$ the dispersion instead exhibits a single minimum  at $p_x=0$. It is also worth discussing the behavior of the effective mass $1/m^*= d^2 \epsilon/dp^2_x$ of particles moving along the $x$-direction. Near the minima one finds \cite{Zhai2013}
\begin{equation}
\frac{m}{m^*}=1- \left(\frac{\hbar \Omega}{4E_r}\right)^2  \;\mathrm{for} \; \hbar \Omega < 4E_r \;\; \mathrm{and} \;\;
\frac{m}{m^*}=1- \frac{4E_r}{\hbar \Omega} \;\mathrm{for} \; \hbar \Omega > 4E_r
\label{m*+-} 
\end{equation}
The effective mass exhibits a divergent behavior at $\hbar\Omega=4E_r$, when the double well structure disappears and the dispersion takes a $p^4_x$ law near the minimum. 
It is also worth noticing that  for $\hbar \Omega < 4 E_r$  the effective mass can get negative values  when one moves away from the minimum, because of the change in the curvature of the function $\epsilon(p_x)$.  This effect is responsible for important nonlinear instabilities that were observed  in the center of mass oscillation   \cite{Zhang2012}  as well as in the expansion of the gas, following the release of the trap \cite{Khamehchi2017}. 
	
\subsection{Quantum Phases and the role of interactions}
\label{sec:phasesSOC}
	
The role of spin-orbit coupling in weakly interacting Bose-Einstein condensates can be properly described employing the mean field energy functional (\ref{Energy}), providing a natural generalization of  Gross-Pitaevskii theory for two coupled Bose-Einstein condensed gases. In the previous Subsection we have already discussed how the value of the Raman coupling $\Omega$ changes in a deep way the structure of the single particle states.  More importantly, interactions  are responsible for the emergence of a new quantum phase, the so called stripe or supersolid phase, which has attracted much interest in the recent literature.  Interactions modify the conditions for the values of $\Omega$ to make the various phases energetically favourable. Choosing a vanishing detuning $\delta=0$   one  identifies the following quantum phases  (see, e.g., \cite{Li2015review}):
	
{\bf Zero Momentum (ZM) phase}. For $\hbar \Omega > 4E_ r - 2g_{ss}n$
the system occupies the $p_x=0$ single particle state. This phase is the analogous of the paramagnetic phase discussed for Rabi coupled BEC. 
It exhibits non trivial features concerning the magnetic, dynamic and superfluid properties. For example, the magnetic susceptibility reads \cite{Li2012}
\begin{equation}
\chi_{ZM}= \frac{2}{\hbar\Omega -(4E_r -2g_{ss}n)},
\label{ChiMZM}
\end{equation}
and exhibits a divergent behavior as one approaches the transition to the Plane Wave phase at $\hbar\Omega = 4E_ r - 2g_{ss}n$ and reduces to \aref{eq:chipara} for $E_r=\hbar^2k_0^2/2m =0$.

{\bf Plane Wave (PW) phase}. As the value of the coupling $\Omega$ is lowered below  $4E_ r - 2g_{ss}n$,  the system enters the so-called plane wave phase, where the gas no longer occupies the $p_x=0$ single particle state, but  states with non vanishing canonical momentum $p_x=\pm\hbar k_1$ that can be written in the form 
	\begin{equation}
	\Psi_+ \equiv  \sqrt{n}\left(\begin{array}{cc} \cos \Theta \\ -\sin \Theta \end{array} \right) e^{ik_1 x}, \;\;\; 
	\Psi_- \equiv  \sqrt{n}\left(\begin{array}{cc} \sin \Theta \\ -\cos \Theta \end{array} \right) e^{-ik_1 x}
	\label{PW+-}
	\end{equation}
\begin{figure}
	\centering
    \includegraphics[width=0.8\textwidth]{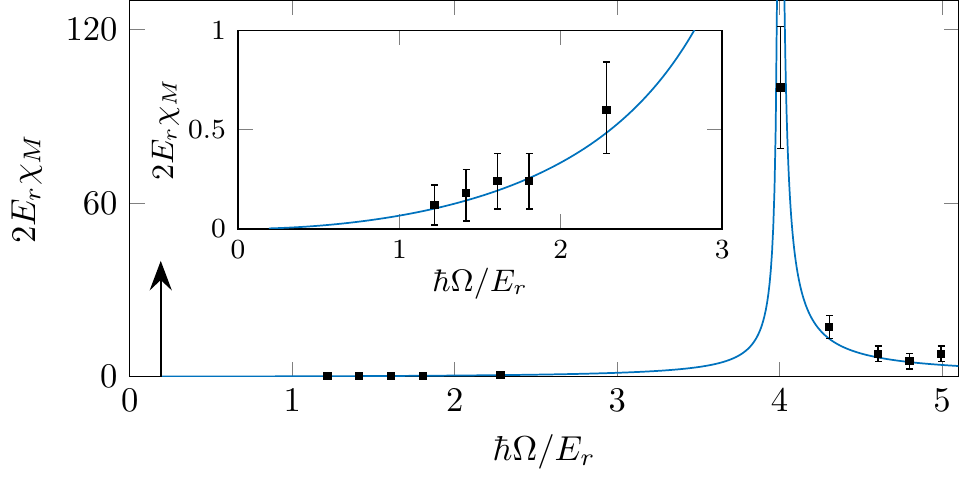}
    \caption{Magnetic susceptibility in the PW and ZM phases. Experimental data are extracted from spin and momentum amplitudes of the dipole oscillation \cite{Zhang2012}. The inset  is a blowup of the small $\hbar\Omega/E_R$ region. Theory is from \cite{Li2012}. The arrow indicates the  transition to the stripe phase. Adapted from \cite{Li2015review}.}
    \label{fig:SOCchiM}
\end{figure}

Minimization of the energy with respect to $k_1$, yields the value $k_1= k_0\sqrt{1- \Omega^2/(4E_r-2g_{ss}n)^2}$ for the wave vector, which renormalizes the non interacting value $k_0$ . The value of the spin polarization is fixed by $k_1$ through the relation $s_z=n\cos(2\Theta)=k_1/k_0$. Notice that $k_1$, and hence  $s_z$  vanishes as one approaches the transition to the zero momentum  phase, corresponding to a second order phase transition.
Since for $\delta=0$ the energy associated with the macroscopic occupation of the two states (\ref{PW+-}) is the same,  the choice between the two configurations is determined by a mechanism of spontaneous symmetry breaking, typical of ferromagnetic configurations. The typical bifurcation exhibited by the canonical momentum $\hbar k_1$ has been explicitly measured by the Spielman team \cite{Lin2011}. The magnetic  polarizability can be easily calculated also in the PW phase, where one finds the result \cite{Li2012}
	\begin{equation}
	\chi_{PW}= \frac{(\hbar \Omega)^2}{(2E_r -g_{ss}n)[4(2E_r-g_{ss}n)^2 -(\hbar \Omega)^2]}
	\label{ChiMPW}
	\end{equation}
	which  diverges as one approaches the transition to the ZM phase. As discussed in the Rabi coupled case, when one approaches the transition   from above or below, the values of $\chi$  differ by a factor  2, reflecting its  second-order nature. If one sets $g_{ss}=0$ in Eqs.(\ref{ChiMZM}) and (\ref{ChiMPW})  the magnetic polarizability turns out to be fixed by  the effective mass introduced in Sect.(\ref{spSOC}) through the relation  $4E_r\chi_M= m/m^*-1$. 
	The magnetic suscptibility has been experimentally extracted \cite{Ji2015} in the ZM and PW phases, through the analysis of the relative amplitude   of the spin and momentum variables  measured after exciting  the dipole oscillation \cite{Li2012}. The extracted values well agree with the theoretical predictions given by Eqs.\ref{ChiMZM} and $\ref{ChiMPW}$ (see Fig.\ref{fig:SOCchiM}).
	Again if $k_0=0$ the polarization $s_z$ and Eq. \ref{ChiMPW} reduce to the Rabi coupled ferromagnetic case, in which case the condition $g_{ss}<0$ is required.
	
	{\bf Stripe Phase (ST)}. In the absence of two-body interactions, the occupation of any combination of the two single particle states (\ref{PW+-})   discussed above is energetically equivalent.  The situation changes  in the presence of interactions and depending on the balance between the density and spin density components of the   interaction terms entering the energy functional (\ref{Energy}), the system prefers to occupy   the above single-particle states either separately, corresponding to the PW configuration, or to occupy the linear combination $\Psi=(\Psi_++ e^{i\phi} \Psi_-)/\sqrt{2}$  . This latter configuration, often called stripe or supersolid configuration, becomes energetically advantageous if the Raman coupling is smaller than the critical value \cite{Ho2011}
	\begin{equation}
	\hbar \Omega_{cr} = 4E_r \sqrt{\frac{2\gamma}{1+2\gamma}} \; ,
	\label{OmegaSP}
	\end{equation}
	where the relevant parameter $\gamma= g_{ss}/g_{dd}$ is fixed by the ratio between the spin and density interaction coupling constants.  Result (\ref{OmegaSP}) holds for positive values of $g_{ss}$ and under the condition $g_{dd}n, g_{ss}n \ll E_r$. The stripe phase has vanishing spin polarization ($s_z =0$) and is characterized by peculiar interference effects between the two components  $\Psi_+$ and $\Psi_-$,  giving rise, for small values of $\Omega$, to density  modulations of the form $	n({\bf r})= \bar{n}\big[ 1 + (\hbar \Omega)/4E_r)\cos(2k_1 x +\phi)\big]$, the actual 
	 position of the interference fringes being  fixed by the value of the phase $\phi$, which results from a mechanism of spontaneous breaking of translational symmetry. For this reason this phase is also called the supersolid phase. Notice also that the space modulation of the density fringes is not uniquely  fixed by $k_0$ since the value of $k_1= k_0\sqrt{1- \Omega^2/(4E_r+g_{dd}n)^2}$ differs from $k_0$, except in the $\Omega \to 0$ limit.  The stripe phase is also characterized by the peculiar behavior 
	 \begin{equation}
	\chi_{SP}= \frac{4(16E_r^2-(\hbar \Omega)^2)}{32E_r^2g_{ss}n-(\hbar \Omega)^2(g_{dd} +2g_{ss})n }
	\label{ChiSP}
	\end{equation}
	 of the magnetic polarizability which exhibits a divergent behavior at the critical point (\ref{OmegaSP}).  Result  (\ref{ChiSP}), similarly to (\ref{OmegaSP}),  holds only if $g_{dd}n, g_{ss}n \ll E_r$.

	In order to increase the contrast of fringes and to reveal more strongly the peculiar effects exhibited by the stripe phase it would be useful to increase the value of $\Omega$. One should however keep in mind that the value of $\Omega$ cannot exceed the critical value (\ref{OmegaSP}) above which the system enters the PW phase. This   value is usually small because in the most familiar  case of alkali atoms the coupling constants relative to the various hyperfine states are very close each other, causing the smallness of $\gamma$.  It has been however  recently shown  \cite{Putra2020} that a rapid  jump in the value of the Raman coupling can provide a way to effectively increase the value of the contrast. In this experiment the jump was in fact slow compared to the gap between the two branches of the SOC dispersion, but fast compared to many body dynamics, which would  bring the system into equilibrium. As a result,  this process simply magnifies  the amplitude of the SOC stripes, making them visible. 
	
	The direct experimental observation of stripes in SOC gases using Bragg spectroscopy  \cite{Li2017,Putra2020} has   provided  one of the first evidences of the long sought phenomenon of supersolidity, where the  spontaneous breaking of gauge symmetry, yielding superfluidity, and of translational invariance, yielding crystallization, co-exist simultaneously.  In \cite{Martone2014} a method was proposed to increase the value of the critical Raman coupling in conditions of thermodynamic equilibrium and  making the   supersolid features of the stripe phase more visible. The proposal is based on the effective reduction of the interspecies coupling constant $g_{\uparrow \downarrow}$, with the consequent increase of $\Omega_{cr}$. It could be  achieved by reducing the spatial overlap between the wave functions of the two spin components, for instance, with the help of a spin-dependent trapping potential separating the two components.

	\subsection{Elementary excitations and Goldstone modes}
	\label{sec:excitationsSOC}
	
\begin{figure}[h!]
	\centering
\includegraphics[width=0.6\textwidth]{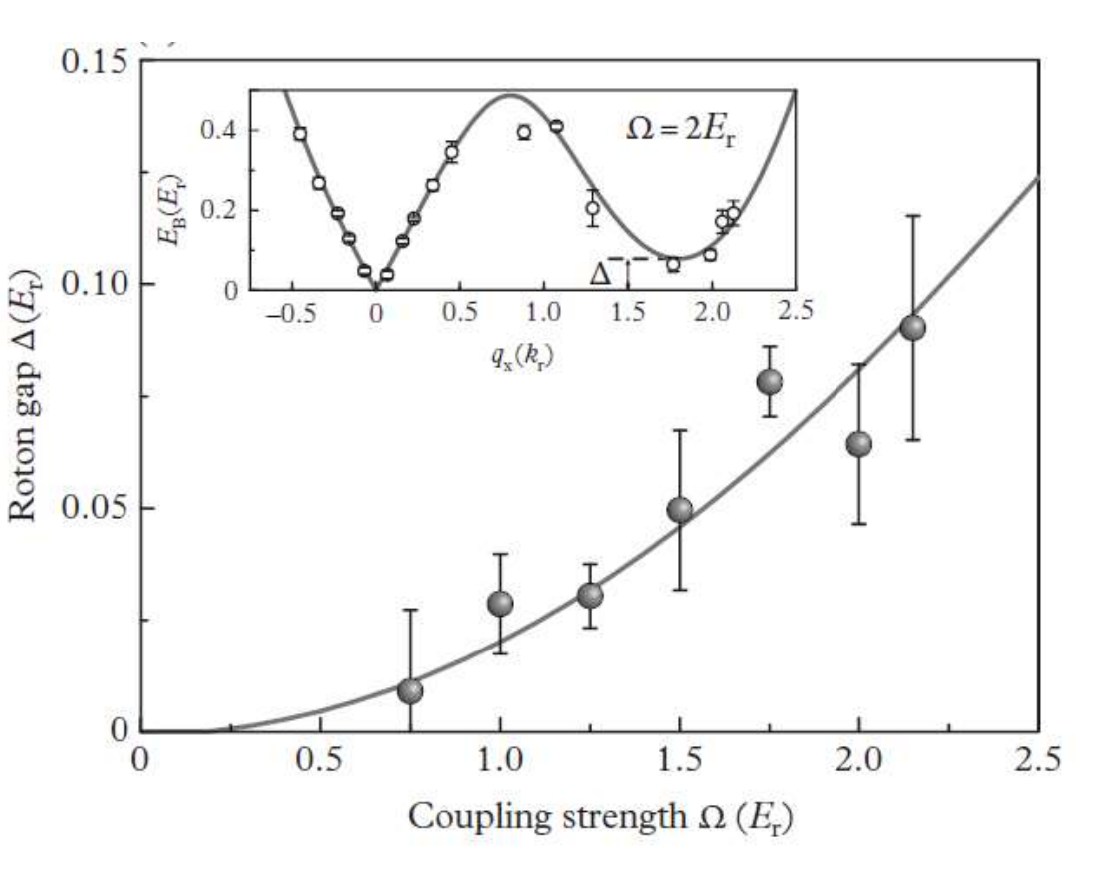}
    \caption{Lower branch of excitations in the plane wave  phase  of a spin-orbit coupled BEC revealing the phonon and roton  excitations. Experimental data are from \cite{Ji2015}.Theory is from \cite{Martone2014}.}
    \label{fig:spectrumSOC}
\end{figure}

	The elementary excitations of a SOC Bose-Einstein condensate can be obtained by solving the  Bogoliubov equations, corresponding to the linearized version of the coupled Gross-Pitaevskii equation. In uniform matter they are classified in terms of the wave vector ${\bf k}$ of the excitation. Similarly to the case of the eigenvalues Eq. (\ref{epsilonssp}) of the single particle Hamiltonian, also the solutions of the Bogoliubov equations exhibit a double band structure, reflecting the spinor nature of the wave function. A typical example of the low branch dispersion is shown in Fig.\ref{fig:spectrumSOC}, obtained in the PW  phase, where the excitation spectrum   measured using Bragg spectroscopy techniques \cite{Ji2015} is reported. 
	Some comments are in order here:  i) the lower branch, exhibits a linear phonon regime for small values of  $q_x$, whose main features can be discussed  using the hydrodynamic formalism; ii) the excitation spectrum violates the symmetry property $\omega(k_x) = \omega(-k_x)$ as a consequence  of the simultaneous violation of parity and time reversal symmetry exhibited by the spin-orbit Hamiltonian; iii) the lower branch in the PW phase exhibits a typical rotonic structure for positive values of $k_x$ ($q_x$ in the figure). Most interestingly the roton gap becomes smaller and smaller as one lowers the Raman coupling strength $\Omega$, approaching the transition to the stripe phase. Since the value of $\Omega_{cr}$ is very small, the transition to the stripe phase  is not visible in the figure. The agreement between experiments and theory is excellent, confirming the validity of the Gross-Pitaevskii mean field approach in the study of the elementary excitations.

	The region of small wave vectors and excitation frequencies can be appropriately described using the hydrodynamic representation in analogy with the description presented in Sec. \ref{sec:josephson} for Rabi coupled BEC's.
	The  hydrodynamic behavior of the system  actually exhibits  very peculiar features \cite{Martone2012}  in the  PW and ZM phases. In uniform (or quasi uniform) matter, where in the large wave length limit  one can neglect quantum pressure effects,  the low frequency oscillations, satisfying the condition $\omega \ll \Omega$ are characterized by the locking of the relative phase: $\phi_r=\phi_\uparrow-\phi_\downarrow =0$ 
	which  reduces the study of the Gross-Pitaveskii equation to the equations for the total density and phase $\phi_d=\phi_\uparrow+\phi_\downarrow$ and the spin-density. These equations, in the linearized limit,  take the form:
	\begin{equation}
	\partial_t \delta n + \frac{1}{2m}{\bf \nabla} \cdot ( n {\bf \nabla} \phi_d)-\frac{k_0}{m}\partial_x \delta s_z=0  \; ,
	\label{HD1}
	\end{equation}
	
	\begin{equation}
	\partial_t\nabla \phi_d+ 2\nabla (g \delta n) =0
	\label{HD2}
	\end{equation}
	and
	\begin{equation}
	-\frac{k_0}{2m}n\partial_x\phi_d+  \frac{\Omega}{2}    \delta s_z =0
	\label{HD3}
	\end{equation} 
	where $\delta n$ and $\delta s_z$ are the fluctuations in the total density and in the spin density taking place during the oscillation.
	For sake of simplicity we have assumed here  $g_{ss}=0$ and  considered the ZM  phase   characterized, at equilibrium, by the vanishing of the phase $\phi$ of the order parameter and of the spin density $s_z$.  The first equation is the equation of continuity, which is deeply affected by spin-orbit coupling, reflecting the fact that the physical current is not simply given by the gradient of the phase as happens in usual superfluids, but contains a crucial spin dependent term. This implies that  even in the density channel the f-sum rule  is not exhausted by the gapless phonon excitation, but can be significantly affected by the higher energy gapped  states, caused by  the Raman coupling. The second equation corresponds to the Euler equation and fixes the time dependence of the phase gradient  of the order parameter. Finally, the third equation follows from the variation of the energy with respect to the spin density and is responsible for the hybridization between the density and spin density degrees of freedom in the propagation of sound \cite{Martone2012}. Notice that if one takes  the $k_0 =\sqrt{2mE_r} \to 0$ limit, corresponding to  the Rabi coupled configurations  discussed in  Sect.\ref{sec:excitationsRabi}, the hybridization disappears and the phonon mode is a  pure density wave. With respect to the hydrodynamic formulation presented in Sect. (\ref{sec:josephson})   for Rabi coupled mixtures, the hydrodynamic equations (\ref{HD1}-\ref{HD3}) cannot describe the dynamics of the relative phase, being applicable only  to the regime of low excitation frequencies, where  $\phi_r$ is locked. 
	
	The linearized equations (\ref{HD1}-\ref{HD3}) can be rewritten in the useful form:
	\begin{equation}
	\partial_t^2 \delta n = \frac{g}{m}[\nabla_\perp\cdot(n \nabla_\perp \delta n)+\frac{m}{m^*}\nabla_x(n \nabla_x \delta n)]
	\label{linHD}
	\end{equation}
	where $n$ is the equilibrium density and  we have introduced the effective mass  $m/m^*=1-\Omega_c/\Omega$. One can show that Eq.  (\ref{linHD}) holds also in the PW phase, taking place for $\Omega < \Omega_{c}$. In this case the effective mass is given by $m/m^*=1-(\Omega/\Omega_c)^2$. 
	In uniform matter, i.e. in the absence of external trapping, Eq.(\ref{linHD}) provides  the phonon  dispersion law  $\omega=ck_x$ along the $x$-direction, with $c^2=g n/m^*$, revealing a strong reduction of the sound velocity in the vicinity of the second order phase transition between the plane wave and the single minimum phase, where the effective mass is much larger than the bare mass (see also Table 1). If one includes the spin term proportional to $g_{ss}$ in the hydrodynamic formalism, one finds that, in the plane wave phase, the sound velocity differs if sound propagates parallel or anti-parallel to the direction fixed by the momentum transfer $\hbar k_0 $ \cite{Li2015review}.  The corresponding velocities $c_+$ and $ c_-$ satisfy the non trivial relation $m\kappa c_+c_-= (1+ 2E_r \chi)^{-1}$, with $1/\kappa = \partial \mu/\partial n$  the inverse compressibility. This relation  reflects the crucial interplay between magnetic effects and the propagation of sound in spin orbit coupled  BEC's. 
	
	The hydrodynamic  equation (\ref{linHD})  allows for analytic solutions also in the presence of harmonic trapping $V({\bf r})= m(\omega_x^2 x^2 +  \omega_y^2 y^2 + \omega_z^2 z^2)$ where the Thomas-Fermi density profile, consistently with the choice $g_{ss}=0$, has the spin-orbit independent form  $n_{eq}= n_0(1-x^2/R_x^2-y^2/R_y^2-z^2/R_z^2)$ and $\omega_x^2R_x^2=\omega_y^2R_y^2=
	\omega_z^2R_z^2$ for positive values of  $n_{eq}$ and $0$ otherwise.  In this case the solution of the hydrodynamic  equations represent an immediate generalization of the results derived in   \cite{Stringari96} in the absence of SOC, with the simple replacement of the trapping frequency $\omega_x$ with $\omega_x\sqrt{m/m^*}$. For example, the frequency of the center of mass oscillation  along the $x$ direcion is expected to be strongly reduced in the vicinity of the transition between the PW and the ZM phase.  This effect  has been experimentally observed in \cite{Zhang2012}.
	
	\begin{figure}
	\centering
\includegraphics[width=0.6\textwidth]{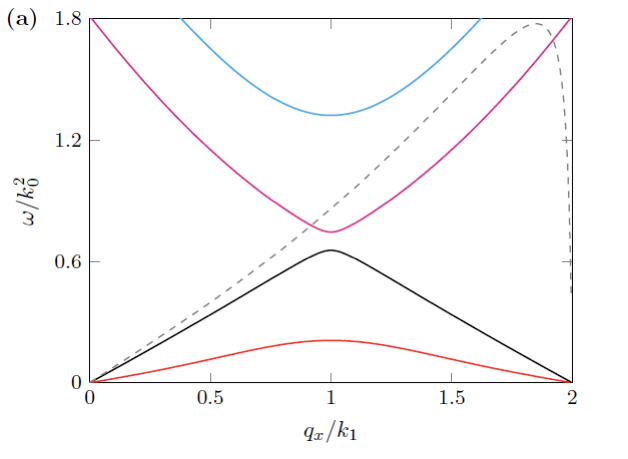}
    \caption{Lowest four excitation bands propagating along the $x$ direction in the stripe phase of a SOC Bose gas ($\hbar=m=1)$. The  lowest gapless branch is the  novel Goldstone mode caused by the breaking of translational invariance.   From \cite{Li2013}.}
    \label{fig:GoldstoneUniform}
\end{figure}

	The dynamic behavior in the stripe phase is particularly interesting because of the novel Goldstone branch introduced by the spontaneous breaking of translational invariance.  
	In uniform matter  this branch approaches, as $\Omega \to 0$,  the spin branch of standard quantum mixtures and is characterized, in the long wave vector limit, by the dispersion $\omega= \sqrt{g_{ss}n/m}k$. 
	The velocity of the novel gapless mode becomes smaller and smaller as one approaches the transition to the plane wave phase and eventually vanishes at the spinodal point.  Both the density and spin density branches propagating along the $x$-direction, exhibit a typical band structure characterized by the Brillouin wave vector $k_1$ (see Fig.\ref{fig:GoldstoneUniform}) and exhibit an important spin-density hybridization effect for finite values of $k_x$. Remarkably, two gapless modes are also predicted to propagate  along the direction parallel to the stripes, where the signal is classified  in terms of the transverse wave vector $k_\perp$. 
	Above the  transition, only the density phonon branch survives, reflecting the fact that only the $U(1)$  symmetry -- associated with the total phase $\phi_d$ of the order parameter -- is spontaneously broken.
	
		\begin{figure}
	\centering
\includegraphics[width=0.7\textwidth]{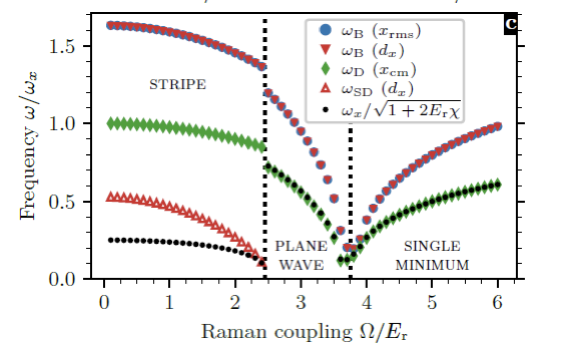}
    \caption{Goldstone modes of a  SOC gas in an axially deformed harmonic trap.  Below the transition between the ST and PW phase one identifies, from top down: axial breathing mode, center of mass oscillation and the novel spin dipole oscillation caused by the spontaneous  breaking of translational symmetry. The black curve corresponds to the upper bound $ \omega_x/\sqrt{1+k_0^2\chi_M}$ to the lowest dipole oscillation, proving the occurrence of a lower frequency mode in the stripe phase. Above the transition the spin dipole mode is fully hybridized to the axial breathing mode  From \cite{Geier2021}}.
    \label{fig:GoldstoneTrap}
\end{figure}
	
	The Goldstone modes  in the stripe phase have been recently theoretically   investigated also in the presence of harmonic trapping \cite{Chen2017,Geier2021}. In this case one finds that, while in the PW and ZM phase the density and spin modes are fully hybridized,  corresponding to the existence of a single gapless branch, in the stripe phase one finds   the emergence of a novel low frequency excitation of spin nature (see Figure \ref{fig:GoldstoneTrap}), whose experimental observation would represent a further  crucial  evidence  of supersolidity. The Figure also shows the result  
	$ \omega_x/\sqrt{1+k_0^2\chi_M}$, which provides  a rigorous upper bound to the lowest mode excited by the dipole operator $x$ \cite{Li2012}. The bound is fixed by the magnetic susceptibility and accurately matches the center of mass frequency both in the PW and in the ZM phases. In the stripe phase the upper bound is instead significantly smaller than the frequency of the collective mode calculated  by solving the time dependent Gross-Pitaevskii equation after the sudden excitation of the  center of mass oscillation. This suggests  the existence of an excitation at lower  frequency which is naturally interpreted as the analog of the  $\omega=0$   Goldstone mode of uniform matter,  corresponding to the translational motion of the stripes.  
	
	
	\subsection{Superfluidity and moment of inertia}
	\label{sec:superfluiditySOC}
	
	Superfluidity is a key feature attracting considerable attention in the theoretical and experimental studies of transport phenomena of  quantum many-body systems at low temperature. It reflects the property that only part (the normal component $\rho_n$) of a system is dragged by the wall of a moving container, the superfluid component $\rho_s =\rho-\rho_n$ being able to move  without friction. Superfluidity exhibits  novel features in spin orbit coupled Bose gases, as a consequence of the violation of Galilean invariance, which affects the usual Landau's criterion for superfludity and the  stability conditions of the superfluid flow \cite{Zhu2012,Ozawa2013,Zhai2013}. 

	A useful definition of the normal density is obtained in terms of the response function of the system to a transverse current perturbation. At zero temperature one can write \cite{Baym1969}:  
	\begin{equation}
	\frac{\rho_n}{\rho} = \frac{1}{N} \lim_{k\to 0}\left[\sum_{n\ne 0} \frac{|\langle 0|J^T_x({\bf k})|n\rangle|^2}{E_n-E_0} +({\bf } \to -{\bf k})\right]
	\label{defBaym}
	\end{equation}	where $J^T_x({\bf k})$ is the transverse  current operator along the $x$-direction (in the following  we will choose the wave vector ${\bf k}$  oriented along the $y$-direction). As  already pointed out in the previous sections, a peculiarity of spin-orbit coupling is that the physical current is not simply given by the canonical contribution, proportional to $p_x$, but  contains an additional spin component. As a consequence, the transverse current operator  takes the form
	  $J^T_x(k_y)  =\sum_k (p_{k,x}-k_0\sigma_{k,z})\exp(ik_yy_k)$. 
		Since the
transverse current operator does not excite the gapless phonon
mode,  of longitudinal nature, the only contribution
to Eq. (\ref{defBaym}) arises from the gapped part of the  spectrum. In both the PW and ZM phases  a single gapped branch is expected to occur. On the other hand, as $k\to 0$ limit, the transverse contribution $|\langle 0|J^T_x(k_y)|n\rangle|^2$  to the sum (\ref{defBaym}), arising from the gapped state, coincides with the corresponding longitudinal one, simply obtained by replacing $\exp(ik_yy)$ with $\exp(ik_xx)$ in the definition of the current. Using sum rule arguments applied to the longitudinal channel it is then possible to show \cite{Ho2011} that the normal density fraction $\rho_n/\rho$ is fixed
by the contribution of the gapped branch to the f-sum rule and that the superfluid density satisfies the important relationship 
\begin{equation}
\frac{\rho_s}{\rho}= m c_+c_n \kappa= \frac{1}{1+k_0^2\chi_M}
\label{c+c-chiM}
\end{equation}
where $c_+$ and $c_-$ are, respectively the velocities of sound propagating parallel and anti-parallel to the $x$-direction of spin-orbit coupling, while $\kappa$ is the compressibility of the gas. In the last equality we have also used the relationship  derive in \cite{Martone2012} between the compressibility $\kappa$, the sound velocities $c_\pm$ and the magnetic susceptibility and   already discussed in Sect.  \ref{sec:excitationsSOC}.   The effects of spin-orbit coupling near the transition between the PW and the ZM  phases is striking, because of the divergent behavior exhibited by $\chi$ (see also Fig.\ref{fig:SOCchiM})  showing that even in configurations of uniform density, the superfluid density of a Bose-Einstein condensed gas at zero temperature is deeply modified  as a consequence of the violation of the Galilean invariance of the Hamiltonian. It is also important to point out that  quantum fluctuations have a negligible consequence  on the depletion of the condensate  \cite{Zhai2013}, thereby revealing that the  superfluid density crucially differs from  Bose-Einstein condensation in these systems.  
The behavior of the superflud density has been also the object of theoretical calculations in the stripe phase \cite{Chen2018,Sanchez2020} where the  occurrence of density modultaions and the consequent emergence of crystal like effects is a further source of reduction of the ratio $\rho_s/\rho$.

A closely related quantity emphasizing the effects of superfluidity is the moment of inertia of a trapped gas. In atomic  quantum gases the moment of inertia  has been  already the object of    theoretical  \cite{Odelin1999,Santo2020} and experimental  \cite{Marago2000,Ferrier2018,Tanzi2021} works  confirming the superfluid behavior of such systems.   It is consequently   interesting to discuss the consequences of spin-orbit coupling. The moment of inertia $\Theta_{inertia}$ around the $z$-axis is defined as the linear response of the system to an external perturbation of the form $H_{pert}=-\Omega_{rot}L_z$, according to the definition $\Theta_{inertia}=\lim_{\Omega \to 0} \langle L_z\rangle)/\Omega{rot}$,
which explicitly reveals  the transverse nature of the response, in analogy with definition \cite{Baym1969} for the normal component of the density.
Deviations of $\Theta_{inertia}$ from the classical rigid value $\Theta_{rig}=m\int d{\bf r}n({\bf r})(x^2 +y^2)$ then point out the consequences of superfluidity. The case of isotropic trapping in the plane of rotation is particularly interesting since in this case the constraint of  irrotationality  on the velocity field imposed by Bose-Einstein condensation in single component configurations implies the vanishing of $\Theta_{inertia}$ \cite{SandroLevBook}. In  SOC gases the situation is  different because the angular momentum contains an additional crucial spin contribution:  $L_z= {\bf r} \times {\bf p} + \hbar k_0y \sigma_z$, thereby 
suggesting that even in the presence of isotropic density configurations in the $x$-$y$ plane, causing the vanishing effect of  the canonical contribution ${\bf r} \times {\bf p}$, the spin term  can provide an important  effect. The calculation of $\Theta_{inertia}$  can be carried out either solving the coupled Gross-Pitaevskii equation in the presence of the  constraint $-\Omega_{rot}L_z$ or using the hydrodynamic equations discussed in the previous Section, with the addition of the term $-\hbar \Omega_{rot} k_0 y$ to the equation (\ref{HD3}) for the spin density. The equation of continuity (\ref{HD1}) is also modified by the new constraint, but with vanishing consequences in the isotropic case.  Making the further simplifying assumption $g_{ss}=0$ the solution of the hydrodynamic equations is analytic \cite{StringariInertia} and in the ZM phase the velocity field  takes the rigid body  like form  ${\bf v}= ( {\bf \Omega}_{rot} \times {\bf r})  \Omega_{cr}/(2\Omega-\Omega_{cr})$ yielding the value 
\begin{equation}
    \Theta_{inertia}= \Theta_{rig}\frac{\Omega_{cr}}{2\Omega-\Omega_{cr}}
\label{vrigid}
\end{equation}
for the moment of inertia. In the PW phase one should simply  replace  the quantity $\Omega/(2 \Omega-\Omega_{cr})$ with 
$\Omega^2/(2\Omega^2_{cr}-\Omega^2$. Remarkably, at the transition between the two phases ($\Omega=\Omega_{cr}$) the moment of inertia takes the rigid value, consistently with the result (\ref{c+c-chiM}) discussed above for the superfluid density, which exactly vanishes at the transition. The above results are confirmed by the numerical solution of the Gross-Pitaevskii equation \cite{Qu2018} and suggest that the inclusion of the detuning $y\sigma_z$   could be actually  employed to provide the experimental  measurement of the moment of inertia in  configurations characterized by isotropic confinement  in the $x$-$y$ plane. 

 For  values of $\Omega_{rot}$ larger than a critical value,   the Gross-Pitaevskii equation  reveals the existence of  an energetic instability, resulting  in the formation of quantized vortices.  This confirms the efficient role played by the inclusion of the $y$-dependence of the detuning to generate non trivial rotational effects, as experimentally proven in the seminal paper \cite{Lin2009} and theoretically discussed in \cite{Galitski2011,Qu2018}.

\section{Conclusions and perspectives}
	
In this paper we have reviewed some key features exhibited by coherently coupled quantum mixtures of Bose-Einstein condensates, providing a combined discussion of Rabi and spin-orbit configurations in $S= 1/2$ spinor mixtures. The emerging scenario emphasizes the rich variety of phenomena exhibited by these systems, including new quantum phases, intriguing  features  of the elementary excitations and of the Goldstone modes as well as  challenging  phenomena, like the internal  Josephson effect, novel solitonic configurations and supersolidity.  The discussion of these phenomena has explicitly  pointed out the crucial role played by the symmetries of the underlying Hamiltonians.  It has also  shown that the  theoretical predictions  and the comparison with the available experiments carried out  in mixtures of ultra cold atomic gases, confirm that mean field formalism, based on the use of Gross-Pitaevskii theory, is a useful starting point for the understanding of the main features exhibited by these systems. At the same time a series of important questions still remains to be explored and understood both from a theoretical and experimental point of view.  Several  specific questions have been discussed in the various sections of the paper. More general open issues, not discussed in this review,     include, among others,  the study of  thermal effects and the interplay between  quantum and thermodynamic  phase transitions,  the crucial role  of quantum fluctuations  in lower dimensions and the novel features exhibited by spinor quantum mixtures with $S\ge 1$. 
	
\section*{Acknowledgement}
We acknowledge  systematic and useful discussions with the members of the experimental and theory groups of the INO-CNR BEC Center. Long standing collaborations on spin orbit coupled gases with Yun Li and Giovanni Martone  and on Bose mixtures with Marta Abad and Albert Gallemì  are warmly acknowledged. This work has been supported by
Provincia Autonoma di Trento, INFN-TIFPA
under the project FIS$\hbar$ and
from the Italian MIUR under the PRIN2017 project
CEnTraL. 
	
	\bibliography{qmixbib}
	\bibliographystyle{ar-style4}
\end{document}